\DocumentMetadata{}
\documentclass[sigconf]{acmart}
\renewcommand\footnotetextcopyrightpermission[1]{} 
\usepackage{orcidlink}
\usepackage{array}
\usepackage{graphicx}
\usepackage{caption}
\usepackage{subcaption}
\usepackage{multirow} 
\usepackage{algorithm}
\usepackage{placeins} 
\usepackage[noend]{algpseudocode}
\usepackage[dvipsnames]{xcolor} 
\usepackage{booktabs}      
\usepackage{lineno}

\AtBeginDocument{%
  \providecommand\BibTeX{{%
    \normalfont B\kern-0.5em{\scshape i\kern-0.25em b}\kern-0.8em\TeX}}}

\setcopyright{none}
\acmDOI{}

\begin{document}

\pagestyle{plain}

\title{PISA: An AI Pipeline for Interpretable-by-design Survival Analysis Providing Multiple Complexity–Accuracy Trade-off Models}

\author{%
\begin{tabular}{ccc}
Thalea Schlender$^{a}$ \orcidlink{0000-0001-7385-112X} &
Catharina J.A. Romme$^{b,c}$ \orcidlink{0009-0001-5815-0306} &
Yvette M. van der Linden$^{a}$ \orcidlink{0000-0002-9003-6124} \\[0.5em]
Luc R.C.W. van Lonkhuijzen$^{b,c}$ \orcidlink{0000-0001-6905-2704} &
Peter A.N. Bosman$^{d}$ \orcidlink{0000-0002-4186-6666} &
Tanja Alderliesten$^{a}$ \orcidlink{0000-0003-4261-7511} \\[0.5em]
\end{tabular}
}
\affiliation{%
  \institution{$^{a}$ Leiden University Medical Center,  Leiden, The Netherlands}
  \country{}
}
\affiliation{%
  \institution{$^{b}$ Amsterdam University Medical Centers location Vrije Universiteit Amsterdam,  Amsterdam, The Netherlands}
  \country{}
}
\affiliation{%
  \institution{$^{c}$ Cancer Center Amsterdam, Cancer Treatment and Quality of Life,  Amsterdam, The Netherlands}
  \country{}
}
\affiliation{%
  \institution{$^{d}$ Centrum Wiskunde \& Informatica,  Amsterdam, The Netherlands}
  \country{}
}



\begin{abstract}

Survival analysis is central to clinical research, informing patient prognoses, guiding treatment decisions, and optimising resource allocation. Accurate time-to-event predictions not only improve quality of life but also reveal risk factors that shape clinical practice. For these models to be relevant in healthcare, interpretability is critical: predictions must be traceable to patient-specific characteristics, and risk factors should be identifiable to generate actionable insights for both clinicians and researchers. Traditional survival models often fail to capture non-linear interactions, while modern deep learning approaches, though powerful, are limited by poor interpretability.

We propose a \textbf{P}ipeline for \textbf{I}nterpretable \textbf{S}urvival \textbf{A}nalysis (\textbf{PISA}) - a pipeline that provides multiple survival analysis models that trade off complexity and performance. Using multiple-feature, multi-objective feature engineering, PISA transforms patient characteristics and time-to-event data into multiple survival analysis models, providing valuable insights into the survival prediction task. Crucially, every model is converted into simple patient stratification flowcharts supported by Kaplan–Meier curves, whilst not compromising on performance. While PISA is model-agnostic, we illustrate its flexibility through applications of Cox regression and shallow survival trees, the latter avoiding proportional hazards assumptions.

Applied to two clinical benchmark datasets, PISA produced interpretable survival models and intuitive stratification flowcharts whilst achieving state-of-the-art performances. Revisiting a prior departmental study further demonstrated its capacity to automate survival analysis workflows in real-world clinical research. 
\end{abstract}


\maketitle

\thispagestyle{empty}

\section{Introduction}


Survival analysis is a statistical approach used to estimate the time until the occurrence of a specific event, such as an infection, disease recurrence, or death. It is specifically designed to appropriately account for censored patients — patients for whom informative but only partial data is available, for instance, due to loss to follow-up or constraints in study duration. In healthcare, survival analysis is crucial not only for predicting individual patient trajectories but also for guiding clinical decision-making. Accurate estimates of time-to-event outcomes enable clinicians to tailor treatment plans, improve patient quality of life, and allocate healthcare resources more efficiently. Moreover, prior knowledge of risk factors and, thus, predisposed patients is invaluable information to guide clinical practice. 


One of the most prominent traditional survival analysis techniques is the Cox proportional hazard regression \cite{cox1972regression}. This semi-parametric model imposes the proportional hazards assumption, i.e. it is assumed that the risk per patient remains constant over time. Further, it assumes a non-parametric base hazard that is only affected by time, multiplied by the linear combination of patient variables and their respective hazard ratios. Hence, no non-linear interactions between the input features can be captured. 
With this, the knowledge discovery of Cox regression models is very limited, as only original input features can be identified as risk factors. This can be (overly) simplistic and may not be able to fully describe the potentially more complicated nature of (combinations of) risk factors that play an important role in the chances of a specific event occurring.



To overcome these limitations, deep survival models have been proposed~\cite{kantidakis2022neural,wiegrebe2024deep}. Deep survival models are able to capture high-order relations in the input data and predict survival, meaning that the neural networks include both automated feature construction as well as a survival estimator. It is important to note that this involves many complex, layered computations, such that these models are not comprehensible by humans.
One such model is DeepSurv, a popular deep learning approach to Cox regression \cite{katzman2018deepsurv}. DeepSurv enhances Cox regression so that it can capture complex interactions of the input features; however, the interpretability and simplicity of the resulting survival analysis model are lost. 


To feasibly use survival analysis models in practice, interpretability is essential. Individual survival predictions should be traceable to patient-specific characteristics, enabling clinicians to understand and trust the basis of each prediction. Moreover, it should be possible to extract meaningful insights, such as relevant risk factors, to inform clinical research.

\begin{figure*}[htbp]
    \centering
    \includegraphics[width=0.95\textwidth]{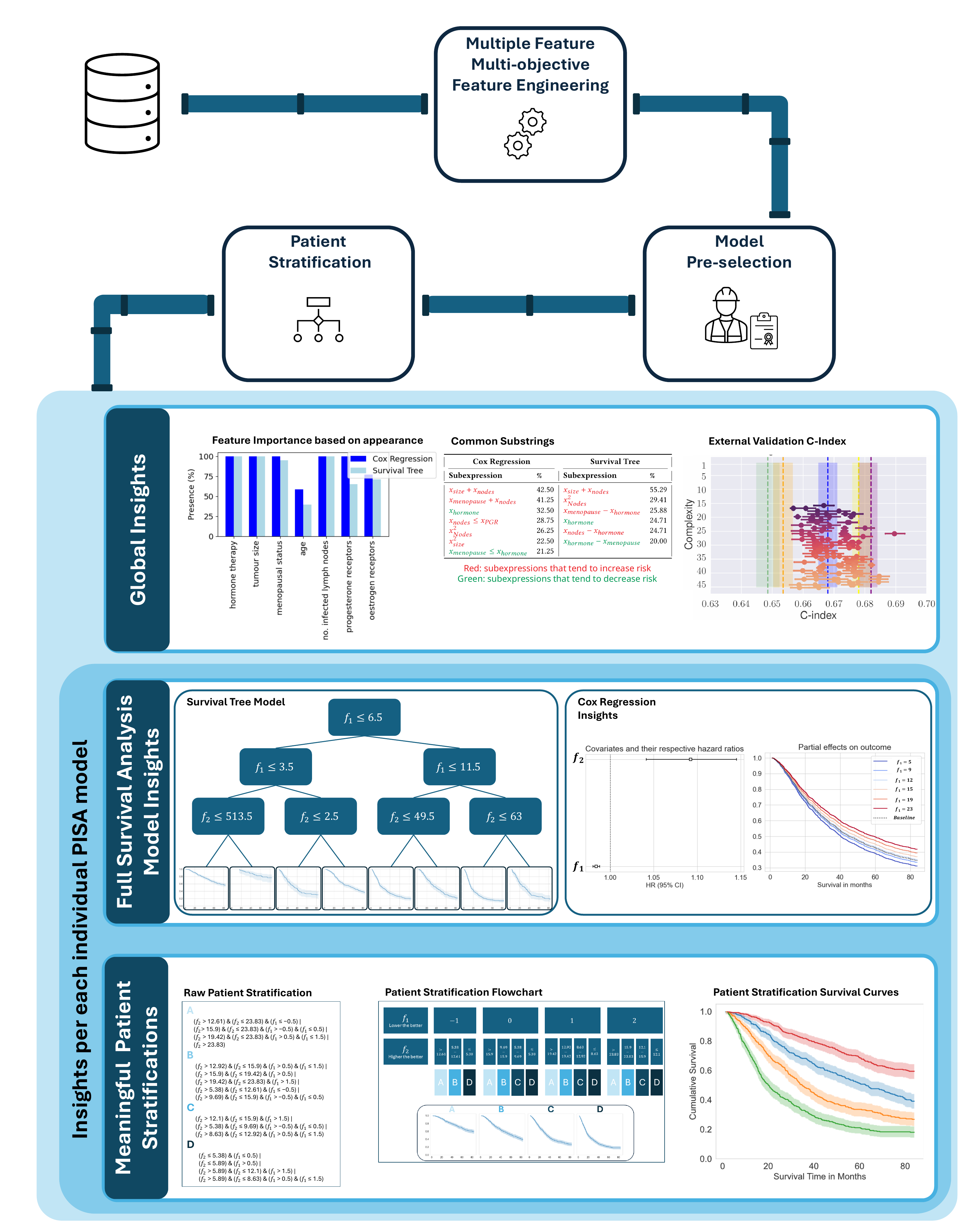}
    \caption{A visual summary of the Pipeline for Interpretable Survival Analysis, \emph{PISA}. From patient characteristics (original input features) and tiem to event data, PISA engineers multiple feature sets consisting of $(f_1,f_2,f_3)$ that trade off how well survival modelling can be done using these features, as well as how complex these features are. The pipeline includes a pre-selection step, after which global insights can be generated from all found feature sets, as well as any individual survival analysis model. Finally, every survival analysis model is automatically transformed into a patient stratification. The visual summary highlights insights that can be derived by using the two different elementary survival analysis models used in this paper: Cox regression and survival trees.}
    \label{fig:Pipeline}
\end{figure*}

Popular survival analysis models that are typically praised for their interpretability are tree-based. Specifically, decision trees have been adapted into survival trees - an example of which is shown in Figure \ref{fig:Pipeline}. These survival trees split the patient population by iteratively determining a split on a single patient input variable that results in two well-separated groups according to their log-rank.
Once each group cannot be split further, the groups use non-parametric estimators, e.g. Kaplan-Meier~\cite{kaplan1958nonparametric}, to estimate the survival distribution. By splitting on multiple single patient input variables, survival trees overcome the Cox regression's limitation of not being able to capture non-linear interactions by performing multiple of these splits, albeit typically making the trees grow large. Another benefit of survival trees is that they avoid the proportional hazards assumption. A survival tree with its intuitive nature is thus a powerful interpretable tool; however, large trees can threaten their interpretability. 

Another approach to capturing complex non-linear interactions, which is also praised for its interpretability, is symbolic regression. Given a set of original input features (i.e patient input variables) and operators (e.g. $+, -, *$ ), symbolic regression describes the task of finding mathematical expressions that make non-linear combinations of original input features to model meaningful relationships and interactions in the data (i.e., in survival analysis: capture relevant risk factors). These expressions (constructed features) are then fed into traditional survival analysis methods. 

To search for these expressions, genetic programming (GP) is a popular technique. Two variants of GP are particularly interesting for survival analysis: multiple-feature and multi-objective GP. The multiple-feature aspect means that multiple features are generated jointly, i.e., at the same time. This enables finding feature sets in which risk factors complement each other to describe survival well. In \cite{scalco2024genetic}, multiple-feature GP is used to engineer multiple features from radiomic input for survival prediction.

The multi-objective aspect means that the features are engineered by optimising for two objectives: how well survival modelling can be done using these features, as well as how complex these features are. Moreover, the search itself proceeds in a multi-objective fashion. This means that the algorithm searches not for one feature set, but a set of feature sets which exhibit different trade-offs between the two objectives. 
In \cite{rovito2025interpretable} the algorithmic potential of using GP to engineer multiple features multi-objectively and embedding these within a Cox model is investigated. Their preliminary work shows that symbolic regression can outperform traditional models, but the results are not validated via bootstrapping or an external validation set. 


Further, their work does not investigate how to use the resulting survival analysis models in practice, nor does it aim to obtain any insights from the models.

Notably, \cite{knottenbelt2025coxkan} proposes CoxKAN: deep survival models using Kolmogorov-Arnold Networks~\cite{liu2024kan}. CoxKAN searches for B-splines for each node in the neural network, so that the network is the sum of learned B-splines for each node. Finally, to add interpretability to the model, they approximate these splines via symbolic regression. However, a limitation is that, as they approximate B-splines rather than fitting the survival problem directly, this may not result in the most suitable mathematical expression. 

Whilst these models have been shown to outperform traditional methods, it is overlooked how survival analysis models are typically used in practice. Specifically, clinicians often rely on meaningful patient stratifications and their respective survival curves for guidance, instead of a full survival analysis model. As such, we propose an automated approach to engineering interpretable survival analysis models that are automatically transformed into patient stratifications without compromising on good performance.

Our \textbf{P}ipeline for \textbf{I}nterpretable \textbf{S}urvival \textbf{A}nalysis (PISA) - as seen in Figure \ref{fig:Pipeline} -, uses multiple-feature multi-objective GP to develop a range of survival analysis models that trade off complexity and performance. 

The performance of a survival analysis model can vary substantially depending on which (subset of the) internal data the model was built with. To mitigate this sample variability as well as the inherent variability in stochastic model construction, it is advisable to repeat the model construction process across different parts of the internal data. Whilst machine learning approaches often construct their modes on different subsets of the internal data, there is little guidance on how to select the final model(s). In multi-objective settings, the number of candidate models increases substantially when this process is repeated. We provide a pre-selection step based on extensive internal validation, allowing fewer models to be presented to the expert while still providing meaningful trade-off options between model complexity and performance.



PISA is an automated pipeline integrating symbolic regression to engineer interpretable feature sets for survival analysis. By generating and validating a spectrum of survival analysis models using feature sets that balance performance and complexity, the pipeline is explicitly designed to prioritise interpretability. To maximise compatibility with clinical settings, each model is converted into a patient stratification tool, visualised through flowcharts, and demonstrated by Kaplan-Meier survival curves. In summary, PISA automatically transforms survival data into relevant patient stratification tools that not only support decision-making but also reveal valuable insights into the underlying risk factors of a survival analysis use case.

Whilst in principle any survival analysis technique may be used within the pipeline, the pipeline is demonstrated by using the Cox regression, as well as survival trees of limited depth. This ensures that we can obtain insights from both the engineered features as well as the resulting survival analysis model. Cox regression, as well as the state-of-the-art methods: DeepSurv, CoxKAN, and the symbolic regression investigation by \cite{rovito2025interpretable} are all bound by the proportional hazards assumption. Whilst we use Cox regression as it is shown to be a good base within the state-of-the-art algorithms, the survival tree of limited depth is particularly interesting due to its interpretability and its absence of the proportional hazards assumption. 

In short, PISA specifically prioritises interpretability and usability of the pipeline, while offering clinicians and medical researchers:

\begin{enumerate}
    \item a selection of different survival analysis models that trade off performance and complexity 
    \item automatically derived meaningful patient stratifications that are validated and summarised via flowcharts. Patient groups are shown via survival curves that are well-separated and achieve good performance.
    \item global insights through analysis of all proposed feature sets, as well as knowledge discovery through each individual interpretable survival analysis model and corresponding patient stratification flowchart
\end{enumerate}

We benchmark our pipeline on two public real-world clinical datasets, which show that using Cox regression and shallow survival trees, the pipeline is capable of generating models that outperform their respective baselines based on the original input features. Further, when the operator set is sufficiently expressive, the pipeline not only achieves state-of-the-art performance but also enables individual models to surpass existing leading methods. When translating the produced survival analysis models to simple patient stratifications, it can be seen that interpretable models can be used without compromising performance. This highlights the pipeline's strong potential for real-world clinical adoption and decision support.

Finally, we revisit a previously conducted study from our department on survival in patients with symptomatic spinal bone metastases. By replicating this study using PISA, we demonstrate the feasibility of applying automation to the clinical research workflow as well as validate the original findings by identifying the same key risk factors.

\section{Results \& Discussion}
\begin{figure*}[htb]
    \centering
    \includegraphics[width=\linewidth]{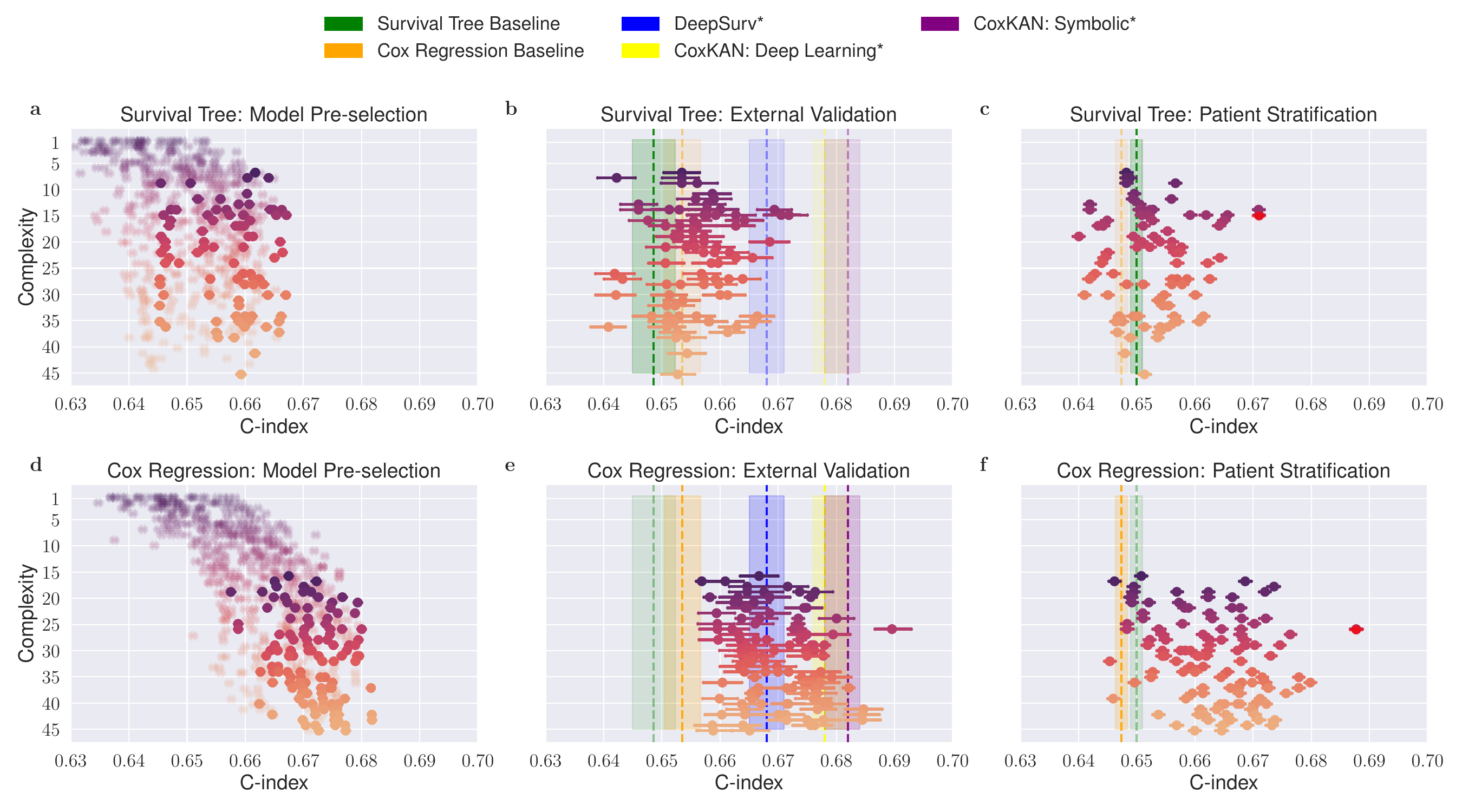}
    \caption{Performance versus complexity plots.
    (a,d) Model pre-selection:  95\% CI of the mean C-index bootstrapped on the internal set. Opaque points indicate the pre-selected models used in PISA.
    (b,e) External validation: Mean C-index of the PISA models on the external validation set with 95\% CI bootstrapping. The mean c-index on the external validation set with 95\% CI bootstrapping of the state-of-the-art deep survival models is shown in both plots. * indicates that the results were taken from their original work. To ease comparison, the baseline models with the same elementary model as the respective PISA plot are opaque, whereas other baselines are translucent.
    (c,f) Patient stratification: Mean C-index of the patient stratification derived from each PISA model on the external validation set with 95\% CI bootstrapping.
    The PISA patient stratifications are compared to the performance of the same stratification procedure applied to the elementary model with original input features. Individual patient stratifications analysed and shown in subsection \ref{indv_mod_GBSG} are highlighted in (c,f) in red. 
    PISA model complexity is shown on the y-axis and additionally encoded using a colour gradient from dark purple (low complexity) to orange (high complexity) to enhance visual clarity.}
    \label{fig:GBSG_Performance}
\end{figure*}

To demonstrate the potential of PISA, we apply and compare our pipeline to traditional and state-of-the-art survival analysis methods on two real-world clinical benchmark datasets - the Rotterdam and German Breast Cancer study group (GBSG)~\cite{foekens2000urokinaseROTTERDAMGBSG,schumacher1994randomizedGBSGGERMAN} and the Study to Understand Prognoses and Preferences for Outcomes and Risks of Treatment (SUPPORT)~\cite{knaus1995support}. The analysis of the latter is shown in the appendix. 
Finally, we revisit a previous study on identifying risk factors for patients with symptomatic spinal bone metastases. The previous study manually analysed the survival data and proposed a patient stratification to guide clinical practice. Our pipeline is applied to this use case to show whether this process can be automated and the study's findings can be confirmed or improved upon. 

All performances are evaluated by calculating the C-index~\cite{harrell1996multivariable}. For the validation of the models, we perform 1000 bootstraps on the external validation set. We provide 95\% confidence intervals of the mean by 1000 bootstrap samples.

\subsection{Survival in patients with node-positive breast cancer}
The GBSG dataset contains data on the overall survival of patients with node-positive breast cancer~\cite{foekens2000urokinaseROTTERDAMGBSG,schumacher1994randomizedGBSGGERMAN}. This real-world clinical dataset has become a popular dataset to benchmark new survival analysis methods with. We obtained the pre-processed GBSG dataset from the DeepSurv~\cite{katzman2018deepsurv} and CoxKAN works~\cite{knottenbelt2025coxkan}, allowing us to replicate their studies exactly.

The subset of 1546 patients of GBSG obtained from the Rotterdam Tumour Bank~\cite{foekens2000urokinaseROTTERDAMGBSG} is used for the derivation of the survival analysis model, whereas the remaining subset of 696 patients obtained from Germany~\cite{schumacher1994randomizedGBSGGERMAN} is used as external validation. The datasets hold information on whether a patient underwent hormonal therapy ($x_{hormone}$), the patients' tumour size ($x_{size}$) stratified into 3 groups (0: 0.5 - 2 mm, 1: 2 -5mm, 2:>5mm), whether a patient is postmenopausal ($x_{menopause}$), the age of a patient ($x_{age}$), the patients' number of positive lymph nodes ($x_{nodes}$), as well as the progesterone ($x_{PGR}$) and oestrogen receptors status ($x_{ER}$).

\subsubsection{Performance Evaluation}
Figure \ref{fig:GBSG_Performance} (b,e) shows external validation performances of the survival analysis models obtained by PISA compared to the survival tree and Cox regression elementary models using the original input features, as well as state-of-the-art (deep-learning based) methods: DeepSurv~\cite{katzman2018deepsurv} and two CoxKAN variants~\cite{knottenbelt2025coxkan}. 

First, it can be seen that the models engineered by our pipeline consistently outperform those produced by their respective baseline methods. As expected, we see a performance difference between the Cox-based and survival-tree-based PISA models. The Cox-based PISA models perform on par with DeepSurv. While this is not the case for CoxKAN, there are still some Cox-based PISA models with equal or better performance. The survival-tree-based PISA models perform on par with the baseline Cox regression, whilst individual models perform as well as DeepSurv. Although the survival-tree-based PISA models generally do not perform as well as the state-of-the-art CoxKAN models, it is important to note that we can achieve Cox regression performance whilst maintaining the superior interpretability of a survival tree with a maximum depth of 3, as well as dropping the proportional hazards assumption.

A further benefit of our pipeline is that it offers multiple survival analysis models to select from, allowing a user to, a posteriori, select a model that fits with their preference. For instance, it is possible to filter the models based on prior knowledge or to trade off the complexity of a model and its performance in a certain way (e.g., prefer very simple models). Further, we can derive insights by analysing the set of engineered models as a whole. For instance, a notion of original input feature importance can be obtained by considering how many models use a certain feature. The feature importance indications for the GBSG use case can be found in the appendix.

Figure \ref{fig:GBSG_Performance} (c,f) shows the external validation performances of the patient stratification models compared to patient stratifications derived from each PISA model, as well as the patient stratification derived using our method from the elementary survival analysis models using the original input features. The different stratifications split the patient cohort into 4-8 groups.

In general, a slight decrease in performance and variance of the 95\% CI of the mean C-index can be noted per individual risk prediction. This can be mainly attributed to the fact that all patients within one group receive the same static risk. Further, performance can decrease with the loss of information due to the chosen stratification boundaries. This can be observed, in particular, for the baseline Cox model, in comparison to the baseline survival tree model, that stratifies patients by design. Even with this slight performance decrease, each individual PISA survival analysis model is transformed into meaningful patient stratifications whilst retaining high performance.
In the next section, we will highlight two stratification models and demonstrate their interpretability. 

\subsubsection{Individual PISA Models}
\label{indv_mod_GBSG}
\paragraph{PISA with Survival Tree elementary model}

\begin{figure*}[htbp]
    \centering
    \begin{subfigure}[b]{0.45\textwidth}
        \centering
        \fbox{\includegraphics[width=\textwidth]{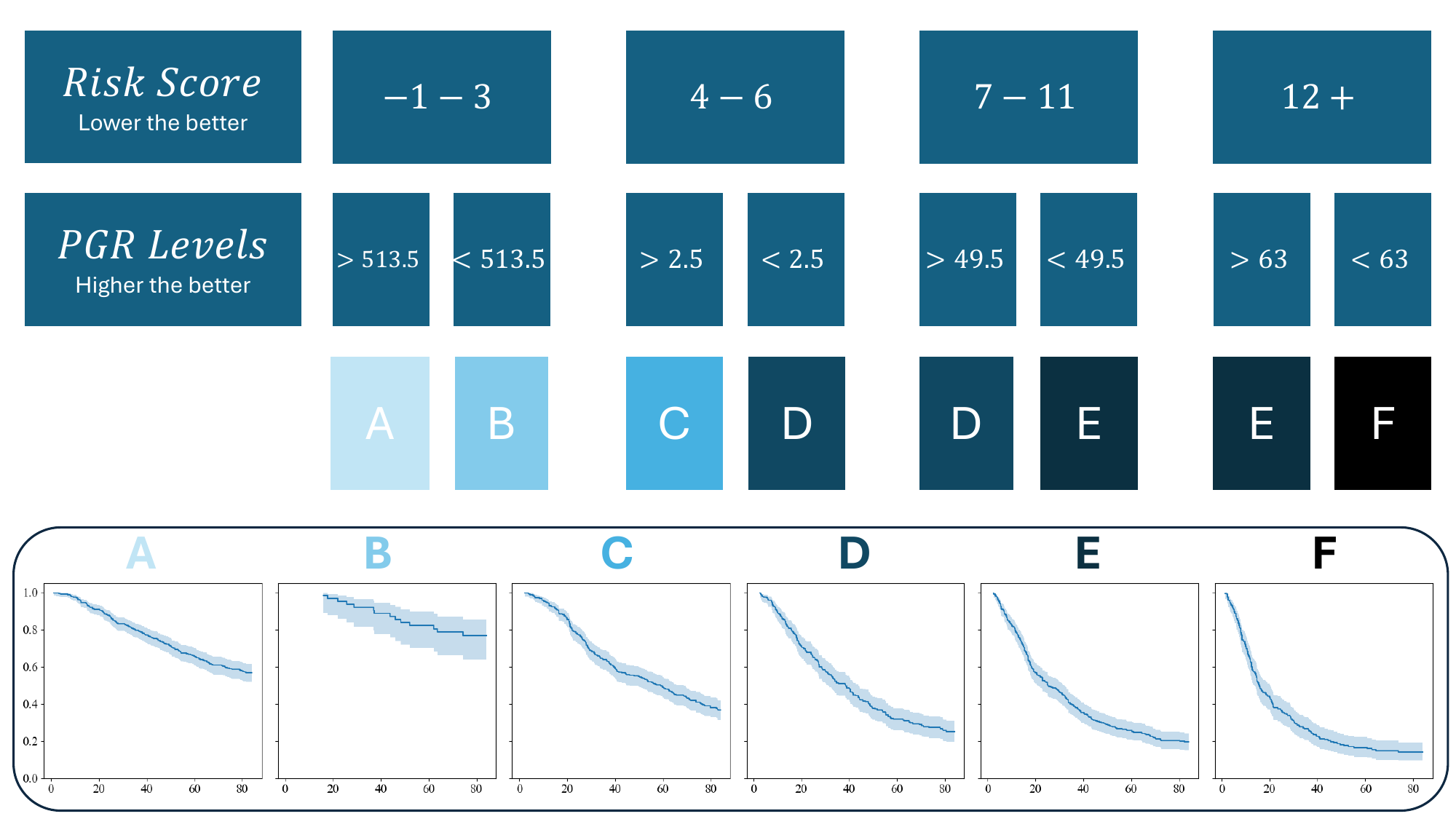}}
        \caption{PISA survival-tree-based model: Flowchart for patient stratification}
        \label{fig:GBSG_ST_STRATsubfig1}
    \end{subfigure}
    \hfill
    \begin{subfigure}[b]{0.45\textwidth}
        \centering
        \fbox{\includegraphics[width=\textwidth]{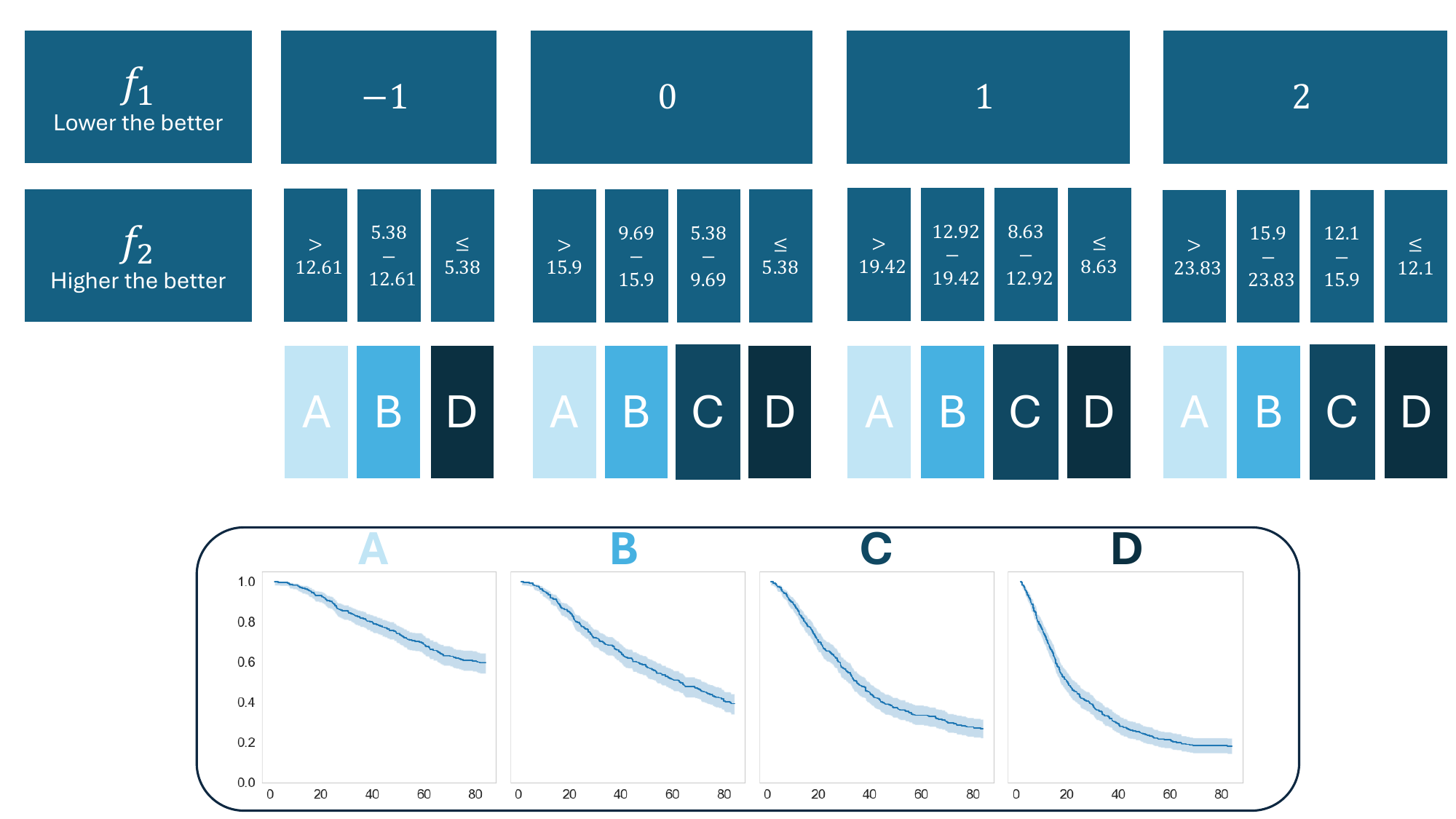}}
        \caption{PISA Cox-based model: Flowchart for patient stratification
        \newline
        }
        \label{fig:GBSG_COX_STRATsubfig1}
    \end{subfigure}
    \begin{subfigure}[b]{0.45\textwidth}
        \centering
        \includegraphics[width=\textwidth]{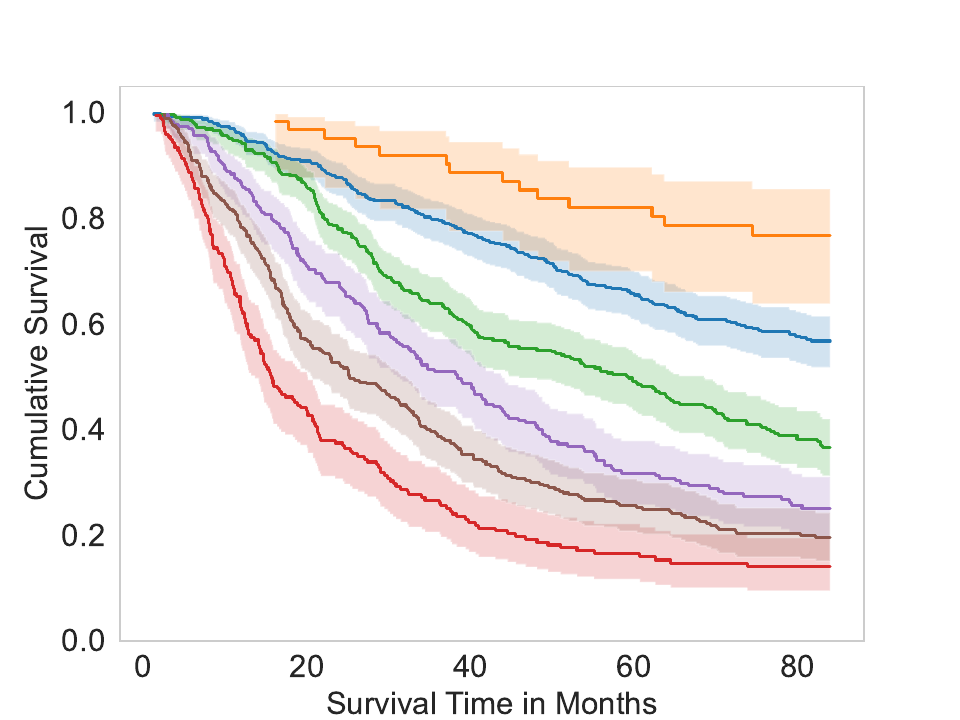} 
        \caption{PISA survival-tree-based model: Survival curves and 95\% confidence interval on the internal dataset}
        \label{fig:GBSG_ST_STRATsubfig2}
    \end{subfigure}
    \hfill    
    \begin{subfigure}[b]{0.45\textwidth}
        \centering
        \includegraphics[width=\textwidth]{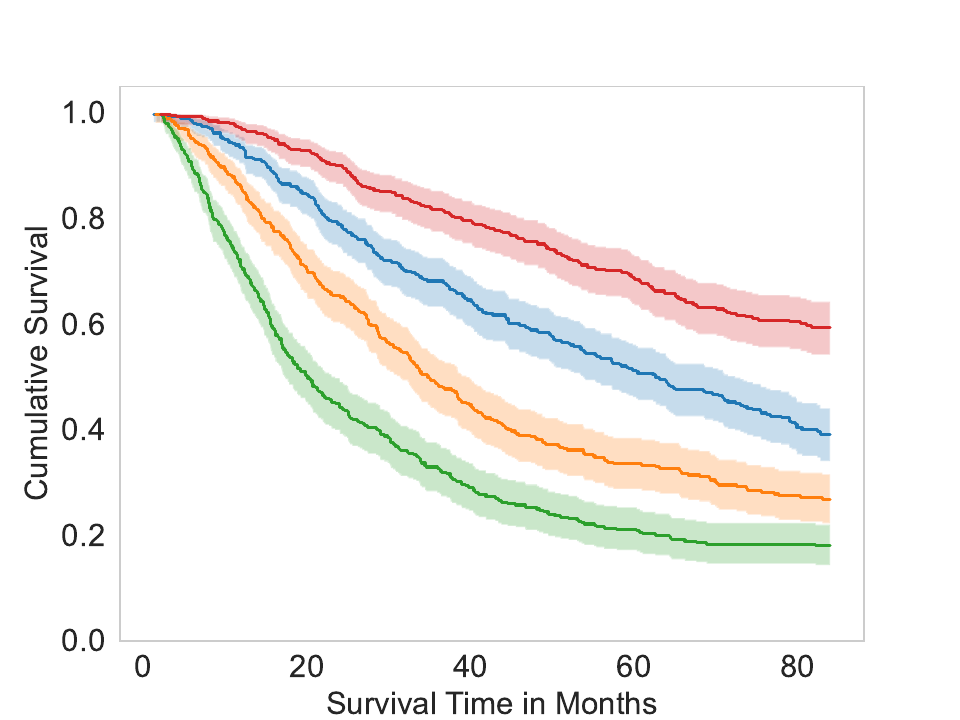} 
        \caption{PISA Cox-based model: Survival curves and confidence interval 95\% on the internal dataset}
        \label{fig:GBSG_COX_STRATsubfig2}
    \end{subfigure}
    \begin{subfigure}[b]{0.45\textwidth}
        \centering
        \includegraphics[width=\textwidth]{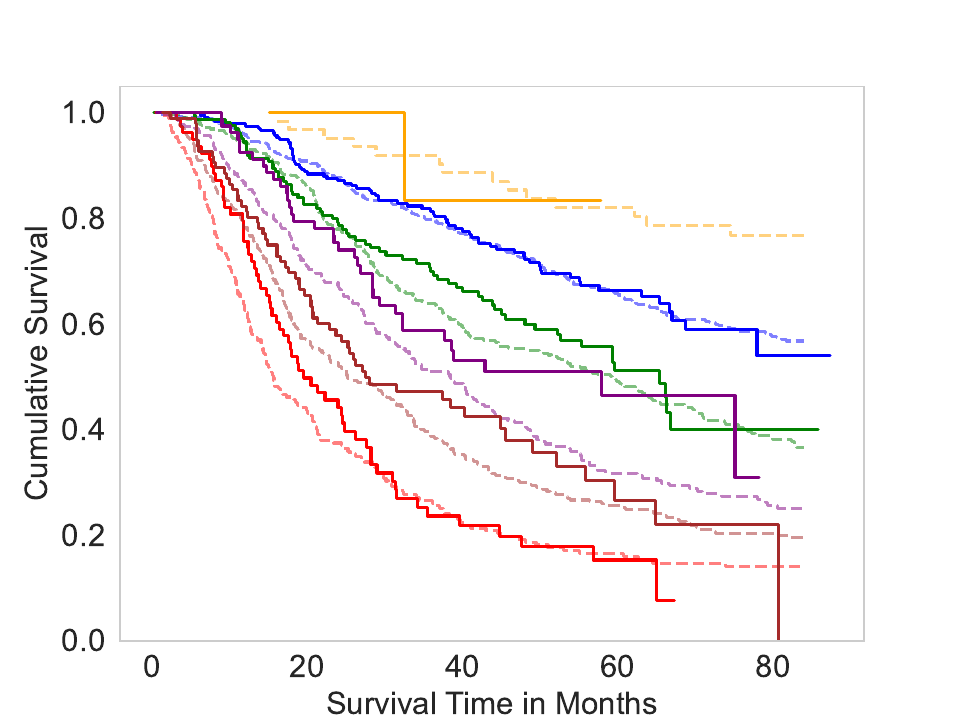} 
        \caption{PISA survival-tree-based model: Survival curves on both the internal (dashed) and the external (solid line) dataset}
        \label{fig:GBSG_ST_STRATsubfig3}
    \end{subfigure}
    \hfill
    \begin{subfigure}[b]{0.45\textwidth}
        \centering
        \includegraphics[width=\textwidth]{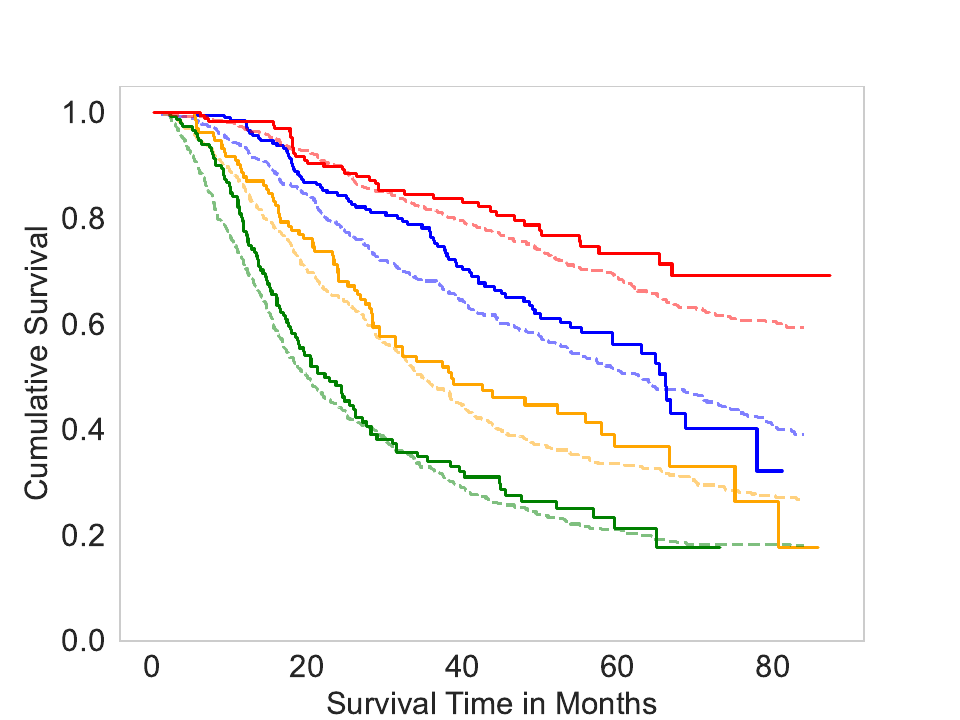} 
        \caption{PISA Cox-based model: Survival curves on both the internal (dashed) and the external (solid line) dataset}
        \label{fig:GBSG_COX_STRATsubfig3}
    \end{subfigure}
    \caption{PISA Stratification Models for GBSG: (a,b) describe flowcharts for patient stratification. (c,d) display survival curves with confidence intervals on the internal dataset, while (e,f) present survival curves on both internal and external datasets. (a,c,e) correspond to a survival-tree-based PISA model; (b,d,f) correspond to a Cox-based PISA model.}
    \label{fig:GBSG_STRAT}
\end{figure*}

To illustrate the patient stratification models, as well as the insights they provide, we show the survival-tree-based patient stratification model here that has the highest external validation performance. The stratification model is based on $f_1$ and $f_2$ given in Equation \ref{Equation:gbsg_st_1} and \ref{Equation:gsbg_st_2} below, achieves a mean C-index of 0.671 on the external validation set, and splits the patients into 6 groups. Figure \ref{fig:GBSG_STRAT} shows the flowchart, as well as the survival curves of the patient groups on both the internal and the external validation set.

 \begin{equation}
    f_1=(x_{ER}\leq x_{age})+x_{size}^2+x_{nodes}+x_{menopause}-2x_{hormone}
\label{Equation:gbsg_st_1}
\end{equation}  
\begin{equation}
    f_2=x_{PGR}
\label{Equation:gsbg_st_2}
\end{equation}

Feature $f_1$ in Equation \ref{Equation:gbsg_st_1} can be seen as a risk score - a higher risk score indicates a higher risk for a patient. Specifically, the risk score linearly combines 5 terms impacting survival. The first term $(x_{ER}\leq x_{age})$ indicates that a higher risk is associated with oestrogen levels (ER) that are lower than or equal to a patient's age. That is, the lower a person's age, the lower the oestrogen level must be to indicate an increased risk. The second term, $x_{size}^2$, indicates higher risk depending on the tumour size, with a steeper increased risk with tumours larger than 5mm, for which $x_{size}=2$. The number of infected nodes $x_{nodes}$ as well as whether a patient has gone through menopause ($x_{menopause}$) also indicate a higher risk. Finally, the last term ($-2x_{hormone}$) shows that if a patient has undergone hormone therapy, the risk decreases. Feature $f_2$ in Equation \ref{Equation:gsbg_st_2} refers to the progesterone levels $x_{PGR}$ only. The flowchart in Figure \ref{fig:GBSG_ST_STRATsubfig1} shows that a higher progesterone level indicates better survival. 

Not only can we confirm the findings of the Cox regression and CoxKAN models of \cite{knottenbelt2025coxkan} in terms of attainable C-index and important features, but we can also contribute a high-performing yet interpretable patient stratification flowchart. When inspecting the derived survival curves on the internal set in Figure \ref{fig:GBSG_ST_STRATsubfig2}, it can be seen that these are well separated. Further, the survival curves on the external validation set in Figure \ref{fig:GBSG_ST_STRATsubfig3} follow the found trends.

\paragraph{PISA with Cox regression elementary model}
The patient stratification based on a PISA-derived Cox model with the highest accuracy achieves a mean C-index of 0.688 on the external validation set, and splits the patients into 4 groups. Based on features $f_1$ and $f_2$ in Equations \ref{Equation:gbsg_cox_1} and \ref{Equation:gbsg_cox_2} below, Figure \ref{fig:GBSG_STRAT} shows the patient stratification flowchart, as well as the survival curves based on the internal and external sets.

\begin{equation}
    f_1=(x_{PGR}\leq x_{menopause}) +(x_{PGR}\leq x_{age}) - x_{hormone}
    \label{Equation:gbsg_cox_1}
\end{equation}
\begin{equation}
    f_2=\frac{2*x_{nodes}+x_{age}+x_{hormone}}{2*x_{menopause}+x_{size}+x_{nodes}}
    \label{Equation:gbsg_cox_2}
\end{equation}

Feature $f_1$ in Equation \ref{Equation:gbsg_cox_1} results in a risk score from $-1$ to $2$, in which a lower score equates to a lower risk. Specifically, this risk score considers 3 terms: first, progesterone levels that are zero if a patient has not yet undergone menopause, and below 1 if a patient has undergone menopause, are found to increase the risk ($x_{PGR}\leq x_{menopause}$). Roughly a fifth of patients in the internal set showcase these extremely low progesterone concentrations. The next term considered in the risk score is whether a patient has a progesterone level lower than or equal to their age ($x_{PGR}\leq x_{age}$). Finally, if a patient has undergone hormone therapy, this lowers their risk ($-x_{hormone}$). 

\begin{figure}
    \centering
    \includegraphics[width=0.95\linewidth]{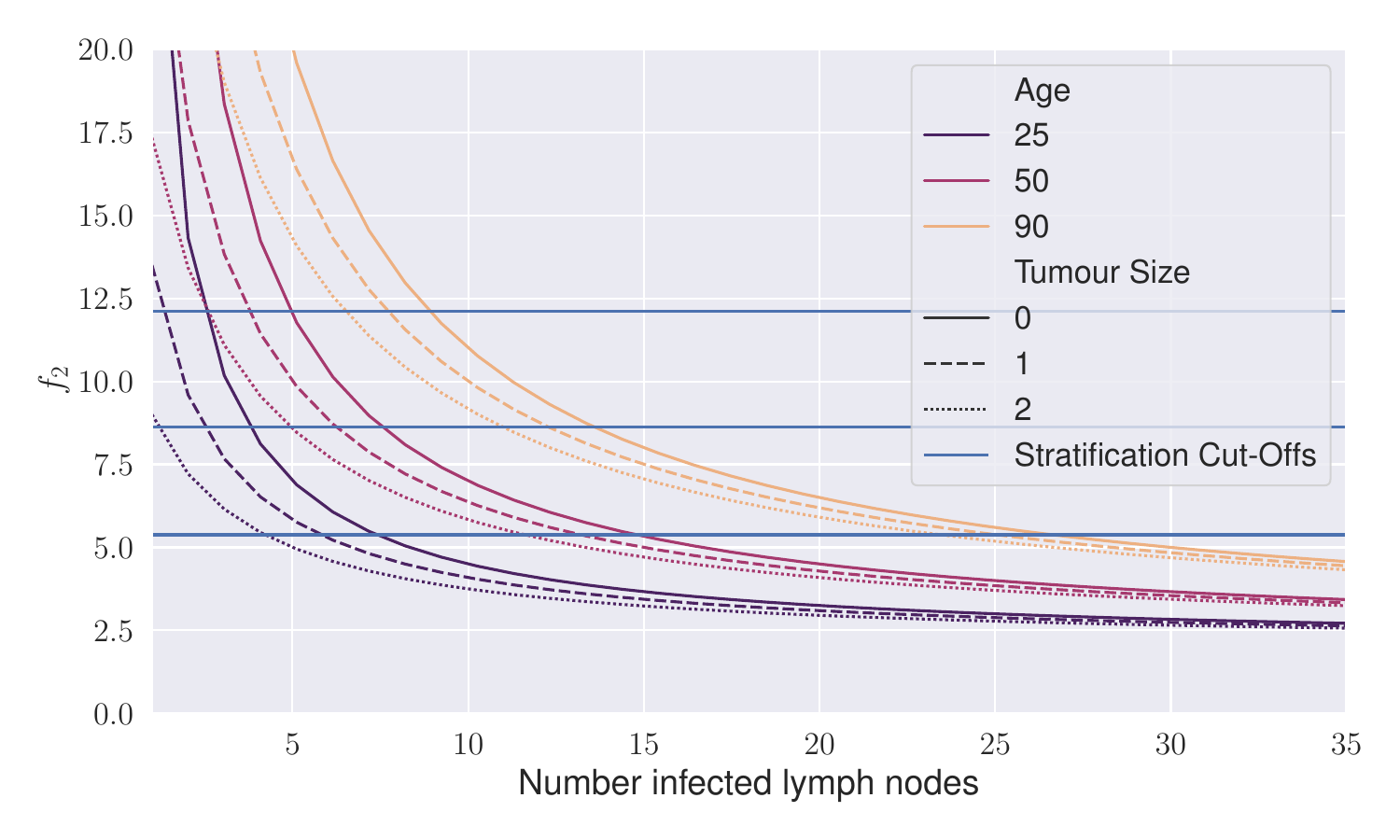}
    \caption{This figure shows how Equation \ref{Equation:gbsg_cox_2} depends on the number of infected lymph nodes, as well as different ages and tumour sizes. The horizontal blue lines refer to the different cutoff points for patient group D (as seen in the flowchart in Figure \ref{fig:GBSG_COX_STRATsubfig1}).}
    \label{fig:GBSG_COX_EQUATION_explained}
\end{figure}

Feature $f_2$ in Equation \ref{Equation:gbsg_cox_2} is slightly more difficult to interpret intuitively. The flowchart in Figure \ref{fig:GBSG_COX_STRATsubfig1} illustrates that lower values are associated with increased patient risk. To aid in interpretation, we show the general tendency in relation to the number of infected lymph nodes, ages, and tumour size in Figure \ref{fig:GBSG_COX_EQUATION_explained}. Specifically, the number of infected lymph nodes is found to have a large impact - a low number of infected nodes produces high feature values indicating low risk for a patient, whereas this risk quickly increases with more infected nodes. Further, a larger tumour size and experienced menopause also increase the risk. Interestingly, the older a patient is, the lower the value. We believe that this could be due to the interplay of the two engineered features together, in which the risk scale already makes cut-off points based on advanced age and the associated postmenopausal status.

The resulting stratification model splits patients into 4 well-defined groups. When inspecting the survival curves of the external set, it can be seen that all survival curves follow the internal curves and are equally well separated. The stratification model splits the patient in meaningful groups with different survival risks, remains interpretable, and outperforms previous state-of-the-art survival models.

\subsection{Survival in patients with symptomatic spinal bone metastases}
Finally, we revisit a previous study from our group on identifying risk factors for poor survival in patients with symptomatic spinal bone metastases. 

The choice of treatment for advanced cancer patients with metastatic disease is crucial to the patient's quality of life. A major factor influencing treatment choices is the trade-off between longer survival and the burden of an extensive treatment, such as increased morbidity and a reduction in the patient's quality of life. As such, an accurate survival prediction is invaluable when comparing different treatment options~\cite{alcorn2020developing,bollen2014prognostic}. 

Specifically, this study aims to predict survival in patients with symptomatic spinal bone metastases that cause a variety of symptoms and typically appear at an end-of-life stage. In \cite{bollen2014prognostic}, different risk factors , as well as a survival prediction model based on patient stratification are proposed by manually selecting features and analysing a Cox regression model built using these features. 

The purpose of PISA is to automate this process, demonstrating that the costs of the labour-intensive manual process can be almost entirely eliminated, potentially find even better models, and offer a wide range of insights into the use case.

The internal dataset contains information on 1043 patients from a single-centre retrospective cohort study of patients treated between January 2001 and December ~\cite{bollen2014prognostic}. For external validation, data was selected from the Dutch Bone Metastases Study~\cite{steenland1999effect} - a national randomised trial. From this study only the 339 patients with symptomatic spinal metastases were included in the validation cohort. It is important to note that the Dutch Bone Metastases Study and the internal data differ significantly. Specifically, the exclusion criteria vary; the internal data focuses on patients with neurological deficit and/or pain, whereas the Dutch Bone Metastases Study explicitly excludes patients with neurological symptoms. Additionally, the Dutch Bone Metastases Study had a significantly shorter follow-up than the internal study (internal dataset has a median follow-up of 6.6 years, whereas the external dataset has a median follow-up of 2.2 years)~\cite{bollen2014prognostic}.

The original input features included in the analysis are: 1) the Karnofsky performance score~\cite{karnofsky1967clinical, schag1984karnofsky} ($x_{kps}$), 2) whether a patient has visceral metastases ($x_{vm}$), 3) whether a patient has brain metastases ($x_{brm}$), 4) the number of bone metastases a patient has (stratified in zero, one or two, three or more) ($x_{bm}$), 5) the number of spinal metastases a patient has (stratified in one, two, three or more) ($x_{sm}$), 6) the location(s) of the spinal metastases (Cervical, Lumbar, Thoracic, Diffuse) ($x_{sm\_C},x_{sm\_L},x_{sm\_T},x_{sm\_D}$) and 7) the clinical profile as derived from \cite{tomita2001surgical} and adapted by \cite{bollen2014prognostic} ($x_{profile\_unfav},x_{profile\_mod},x_{profile\_fav}$).

\begin{figure*}[htb]
    \centering
    \includegraphics[width=\linewidth]{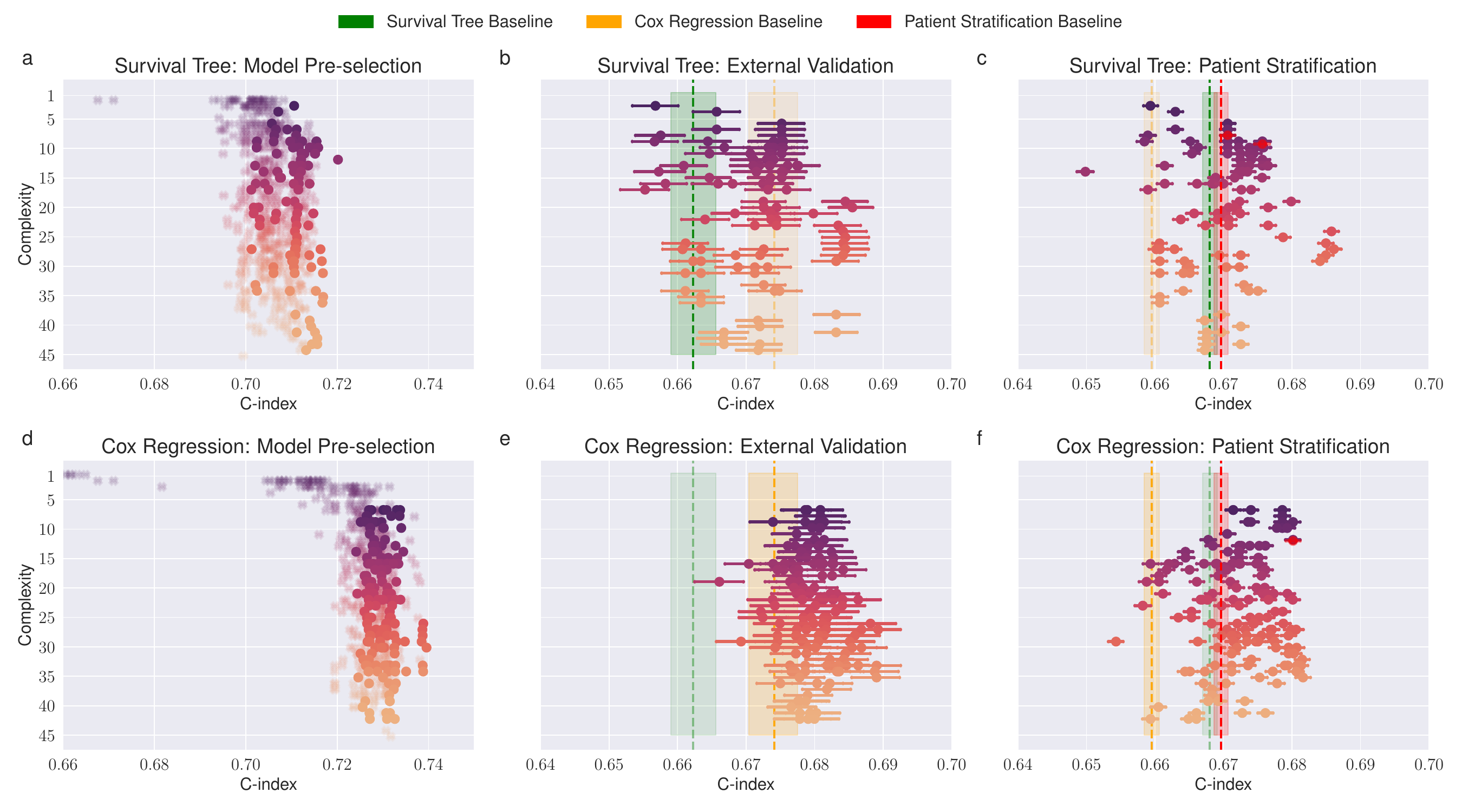}
    \caption{Performance versus complexity plots.
    (a,d) Model pre-selection:  95\% CI of the mean C-index bootstrapped on the internal set. Opaque points indicate the pre-selected models used in PISA.
    (b,e) External validation: Mean C-index of the PISA models on the external validation set with 95\% CI bootstrapping. The mean c-index on the external validation set with 95\% CI bootstrapping of the state-of-the-art deep survival models is shown in both plots. To ease comparison, the baseline models with the same elementary model as the respective PISA plot are opaque, whereas other baselines are translucent.
    (c,f) Patient stratification: Mean C-index of the patient stratification derived from each PISA model on the external validation set with 95\% CI bootstrapping. Individual patient stratifications analysed and shown in subsection \ref{indv_mod} are highlighted in (c,f) in red. 
    The PISA patient stratifications are compared to the performance of the same stratification procedure applied to the elementary model with original input features. Further, the 95\% CI for the mean c-index bootstrapped on the external validation set is given for the proposed patient stratification of \cite{bollen2014prognostic}'s original work.
    PISA model complexity is shown on the y-axis and additionally encoded using a colour gradient from dark purple (low complexity) to orange (high complexity) to enhance visual clarity.}
    \label{fig:PAL_performance}
\end{figure*}
\subsubsection{Performance Evaluation}
\begin{figure}
    \centering
\includegraphics[width=0.95\linewidth]{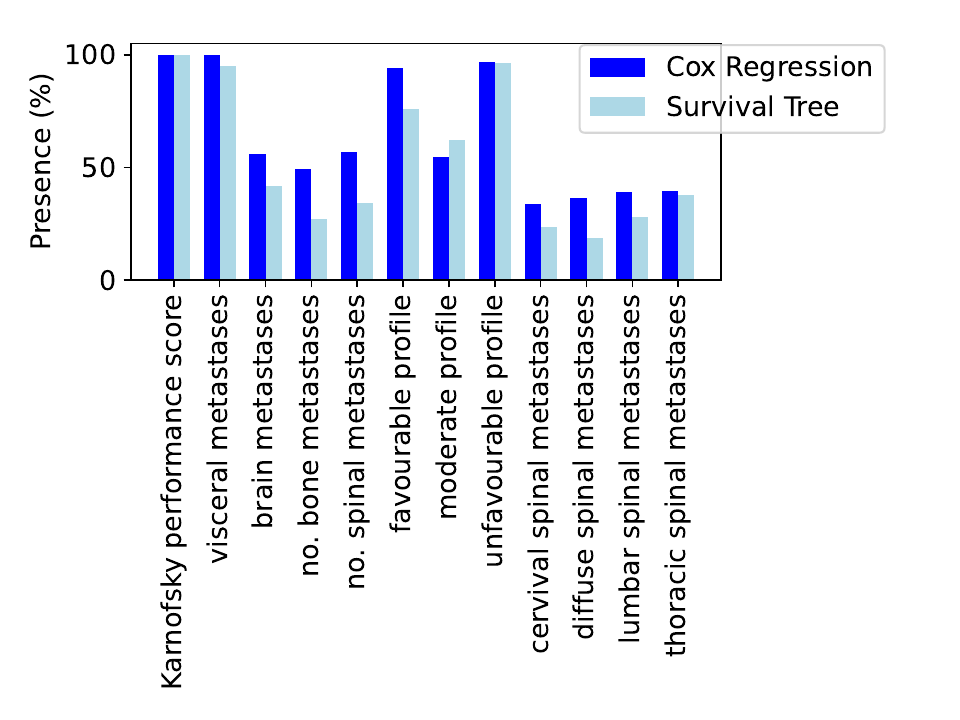}
    \caption{Feature importance scores of the spinal dataset based on their frequency of appearance in the PISA survival analysis models.}
    \label{fig:PAL_Feature_importance}
\end{figure}
Figure \ref{fig:PAL_performance} (b,e) shows the external validation performances of the survival analysis models compared to the survival tree and Cox regression baselines. It can be seen that the PISA survival analysis models using both Cox regression as well as shallow survival trees as elementary models outperform their respective baselines, indicating a better performance. 

To obtain global insights in terms of feature importance indications, we analyse the resulting models. Figure \ref{fig:PAL_Feature_importance} reveals that the Karnofsky score, the presence of visceral metastases and the clinical profile are most important and always appear in the models. Note that it is not necessary to have information on all clinical profiles included in the expressions, as all patients belong to one of the three profiles, and, thus, information on two profiles is sufficient. Further, the original work uses domain expertise to group the occurrence of visceral and brain metastases into one indicator. We chose not to do this in order to test whether the pipeline would also identify both of these as important. Information on brain metastases was included in 56\% and 42\% of the Cox-regression and survival-tree-based PISA survival analysis, respectively. We believe this is due to the low prevalence of brain metastases, present in only 6.74\% of patients in the internal dataset and 2.36\% in the external dataset.

Figure \ref{fig:PAL_performance} (c,f) shows the external validation performances of the patient stratification tools, compared to the patient stratification tool proposed in \cite{bollen2014prognostic}. Again, a slight performance and variance decrease can be detected, but the stratification tools outperform the proposed stratification tool of \cite{bollen2014prognostic} in terms of C-index. However, the original tool splits the patient population into 4 groups, whereas the stratifications from our pipeline propose splits into 5 - 8 groups, which is slightly more complex. When analysing the simplest top performing models further, it can be seen that many models use the same original input features as proposed originally, specifically, the Karnofsky score, clinical profile, and occurring visceral metastases. The original work splits the Karnofsky performance score between 10-70 and 80 -100. In contrast, we find different models using different splits between good and poor Karnofsky scores, as well as changing the splits between good and poor performance based on the other original input features used. We believe this could be the reason for the performance increase.

In the next subsection, two simple stratification tools are shown that trade off performance and complexity. In contrast to the clinical benchmark problems seen so far, the survival times differ drastically between the internal and the external set. 
As such, the survival curves are not compared directly, but are observed to see how well the survival curves are separated. 
Further, the difference in results on the internal and external validation set highlights the need for the best-performing models to perform well on both the internal and external datasets.

\subsubsection{Individual PISA Models}
\label{indv_mod}
To demonstrate the individual models found by the pipeline, we show the simplest but well-performing models here. 

\paragraph{PISA with Survival Tree elementary model}
\begin{equation}
    f_1=(x_{profile\_mod}+x_{vm}+\frac{329.296}{x_{kps}})^2
    \label{Equation:PAL_ST_1}
\end{equation}
\begin{equation}
    f_2=x_{profile\_unfav}
    \label{Equation:PAL_ST_2}
\end{equation}

\begin{figure*}[htbp]
    \centering
    \begin{subfigure}[b]{0.45\textwidth}
        \centering
        \fbox{\includegraphics[width=\textwidth]{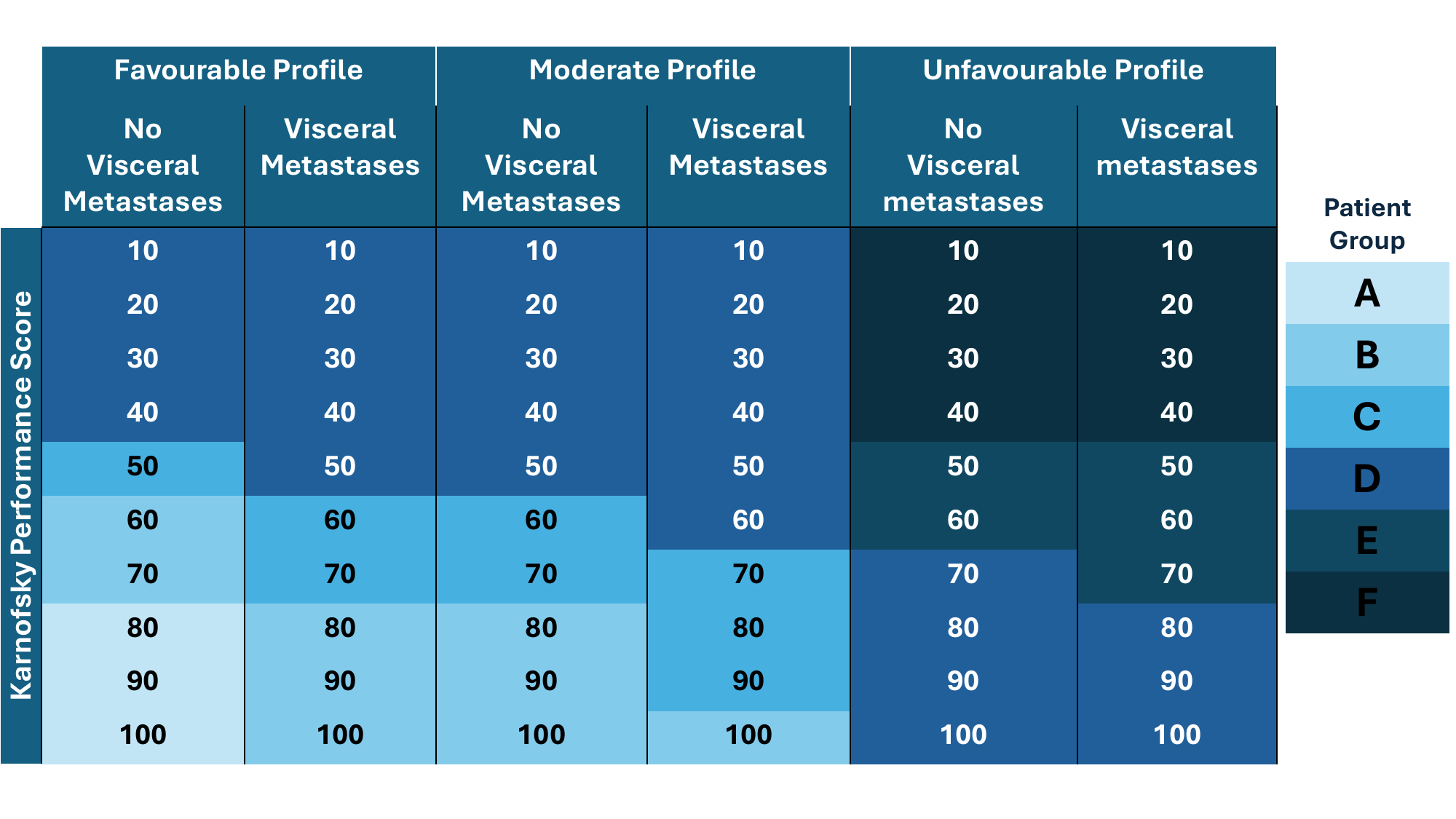}}
        \caption{PISA survival-tree-based model: Flowchart for patient stratification}
        \label{fig:PAL_ST_STRATsubfig1}
    \end{subfigure}
    \hfill
    \begin{subfigure}[b]{0.45\textwidth}
        \centering
        \fbox{\includegraphics[width=\textwidth]{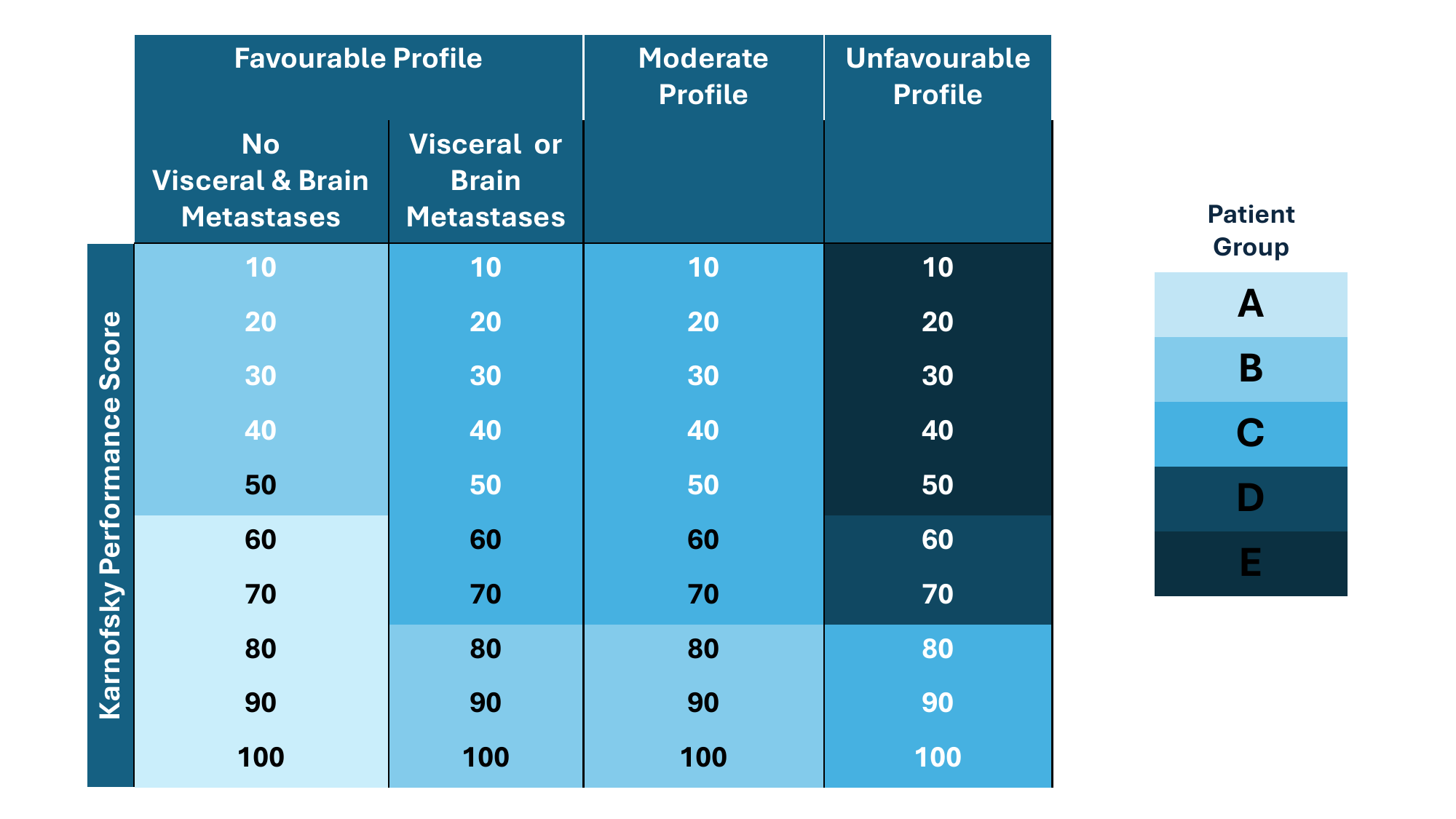}}
        \caption{PISA survival-tree-based model: Flowchart for patient stratification}
        \label{fig:PAL_ST2_STRATsubfig1}
    \end{subfigure}
    \begin{subfigure}[b]{0.45\textwidth}
        \centering
        \includegraphics[width=\textwidth]{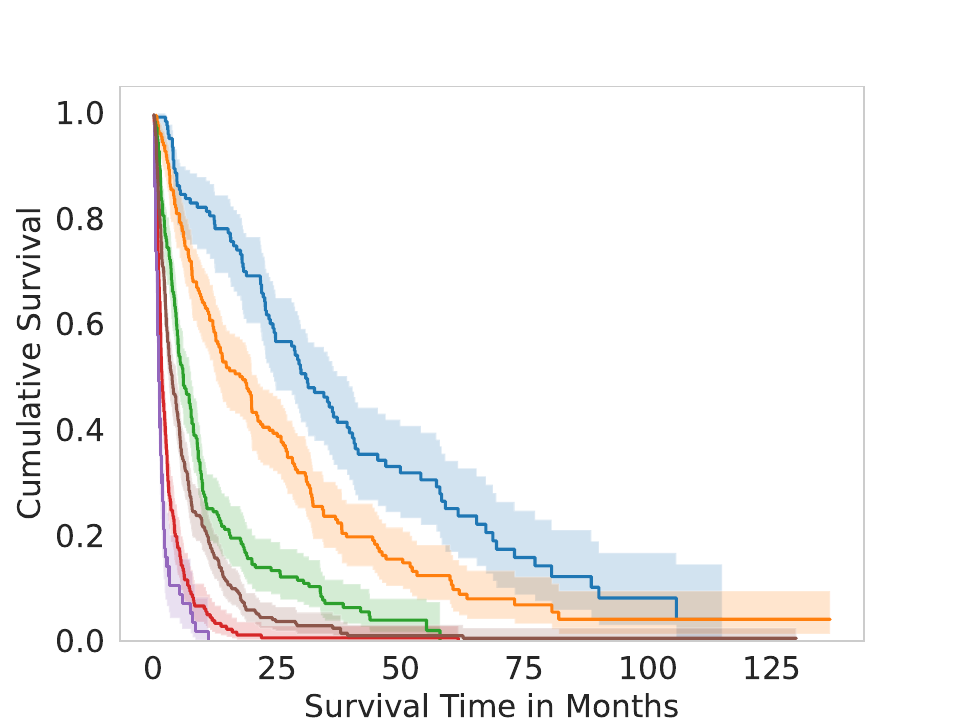} 
        \caption{PISA survival-tree-based model: Survival curves and 95\% confidence interval on the internal dataset}
        \label{fig:PAL_ST_STRATsubfig2}
    \end{subfigure}
    \hfill    
    \begin{subfigure}[b]{0.45\textwidth}
        \centering
        \includegraphics[width=\textwidth]{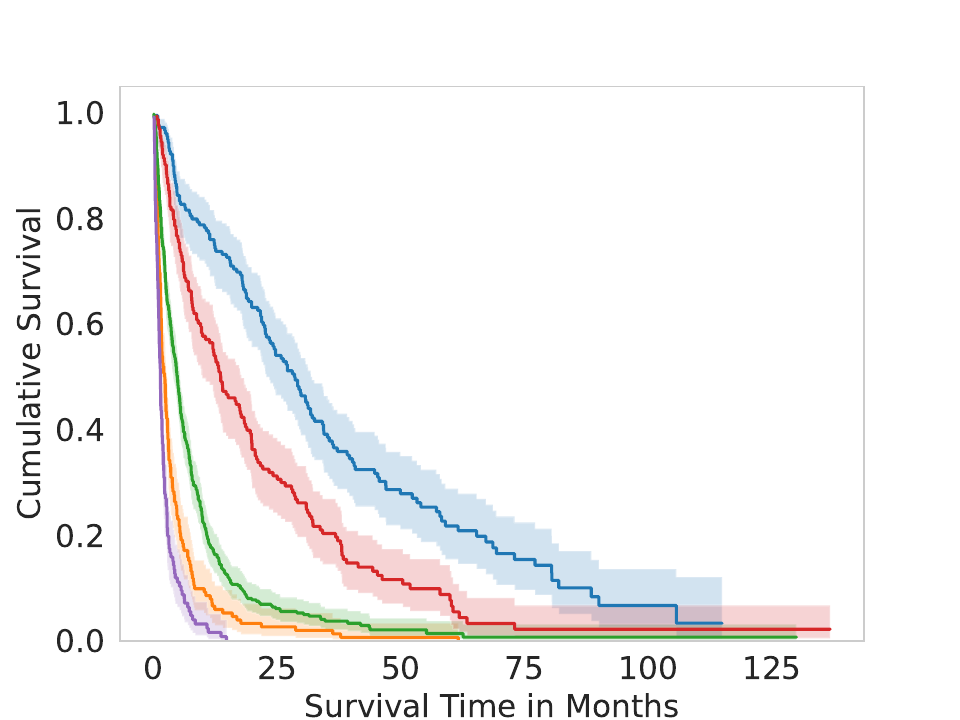} 
        \caption{PISA survival-tree-based model: Survival curves and 95\% confidence interval on the internal dataset}
        \label{fig:PAL_ST2_STRATsubfig2}
    \end{subfigure}
    \begin{subfigure}[b]{0.45\textwidth}
        \centering
        \includegraphics[width=\textwidth]{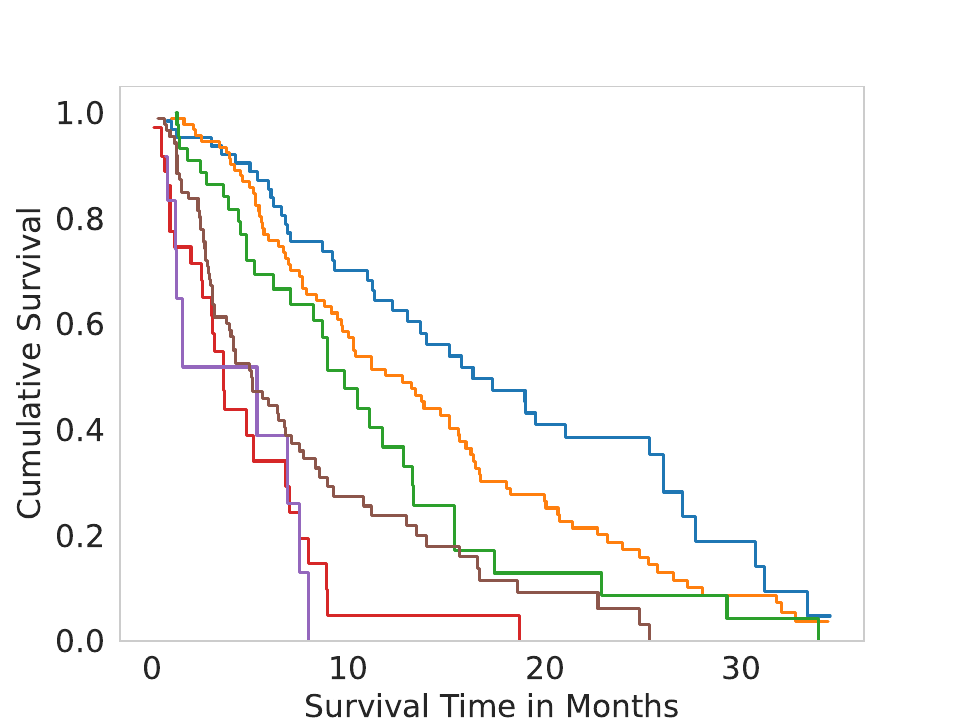} 
        \caption{PISA survival-tree-based model: Survival curves on the external dataset}
        \label{fig:PAL_ST_STRATsubfig3}
    \end{subfigure}
    \hfill
    \begin{subfigure}[b]{0.45\textwidth}
        \centering
        \includegraphics[width=\textwidth]{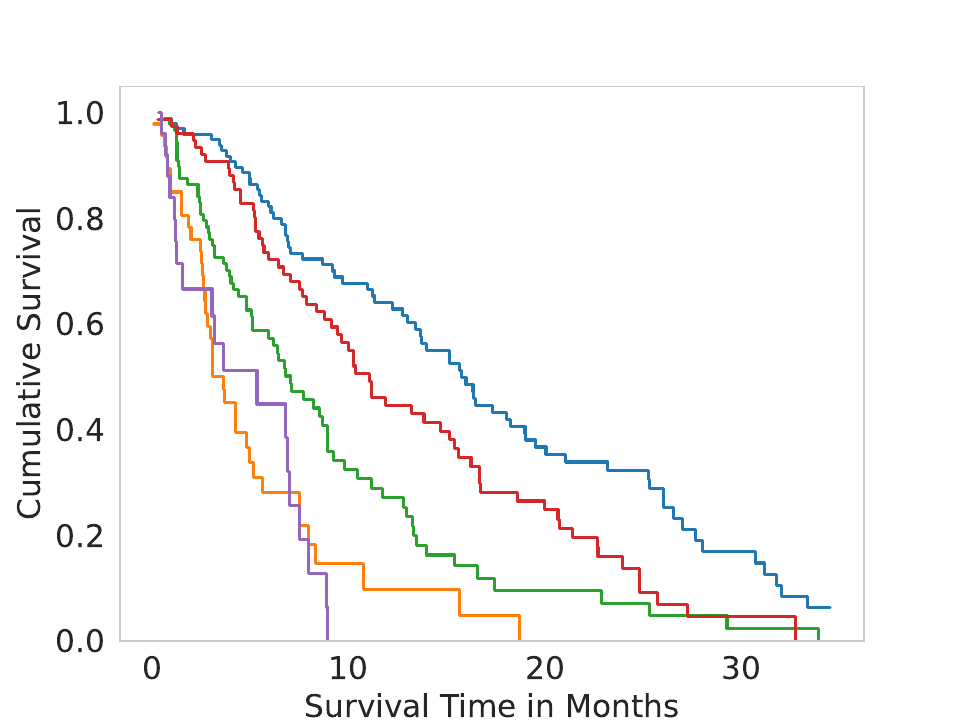} 
        \caption{PISA survival-tree-based model: Survival curves on the external dataset}
        \label{fig:PAL_ST2_STRATsubfig3}
    \end{subfigure}
    \caption{PISA Stratification Models for the spinal metastases use case: (a,b) describe flowcharts for patient stratification, (c,d) show survival curves with 95\% confidence intervals on the internal dataset, while (e,f) show survival curves on the external datasets. (a,c,e) show the patient stratification of a PISA survival-tree-based model from Equations \ref{Equation:PAL_ST_1} and \ref{Equation:PAL_ST_2}, while (b,d,f) show the patient stratification of a PISA survival-tree-based model from Equations \ref{Equation:PAL_ST2_1}, \ref{Equation:PAL_ST2_2} and \ref{Equation:PAL_ST2_3}.}
    \label{fig:PAL_STRAT}
\end{figure*}

Figure \ref{fig:PAL_STRAT} shows the stratification model built on features $f_1$ and $f_2$ in Equations \ref{Equation:PAL_ST_1} and \ref{Equation:PAL_ST_2}, as well as the resulting survival curves. The stratification tool splits the patient population into 6 groups and achieves an external validation C-index of 0.676.

A lower Karnofsky score ($x_{kps}$), unfavourable clinical profile ($x_{profile\_unfav}$), and the presence of visceral metastases ($x_{vm}$) result in a lower survival. The flowchart in Figure \ref{fig:PAL_STRAT} shows how the Karnofsky score ($x_{kps}$) is stratified into different risk groups, according to threshold values and other factors.  In general, the survival curves stratified the two patient populations into groups well. It is noticeable, however, that the patient group $F$ with the worst prognosis overlaps with the other survival curves in the external set. The original work mentions that more patients with a favourable clinical profile entered the external study, and fewer patients in the patient group $F$, making it sensitive for the individual outcomes.

Like the original study, $f_1$ (Equations \ref{Equation:PAL_ST_1}) and $f_2$ (Equation \ref{Equation:PAL_ST_2}) use the clinical profile ($x_{profile\_mod}, x_{profile\_unfav}$), the Karnofsky score ($x_{kps}$), and the presence of visceral metastases to stratify patients, but not include information on the presence of brain metastases ($x_{brm}$). The flowchart outperforms the original patient stratification without this information, potentially signalling that knowledge on the presence of brain metastases is not essential. 

\begin{equation}
    f_1=x_{profile\_mod}+NOT(x_{vm}\ or\ x_{brm})
    \label{Equation:PAL_ST2_1}
\end{equation}
\begin{equation}
    f_2=x_{kps}
    \label{Equation:PAL_ST2_2}
\end{equation}
\begin{equation}
    f_3=x_{profile\_unfav}
    \label{Equation:PAL_ST2_3}
\end{equation}

We highlight, however, that if a clinician wishes to only consider survival analysis models incorporating information on brain metastases, they can filter the final models accordingly. Figure \ref{fig:PAL_STRAT} shows another stratification model using brain metastases ($x_{brm}$) in addition to the previously mentioned original input features. The model is built using features $f_1, f_2$ and $f_3$ in Equations \ref{Equation:PAL_ST2_1}, \ref{Equation:PAL_ST2_2}, and \ref{Equation:PAL_ST2_3}, with which the patient population is stratified into 5 groups. The model achieves an external validation C-index of 0.671, and its survival curves are shown in Figure \ref{fig:PAL_ST2_STRATsubfig2} as well as Figure \ref{fig:PAL_ST2_STRATsubfig3}.

\paragraph{PISA with Cox regression elementary model}
\begin{equation}
    f_1=x_{kps}
    \label{Equation:PAL_COX_1}
\end{equation}
\begin{equation}
    f_2=x_{profile\_unfav}
    \label{Equation:PAL_COX_2}
\end{equation}
\begin{equation}
    f_3=((x_{brm}-x_{profile\_fav})-(~x_{vm}))^2
    \label{Equation:PAL_COX_3}
\end{equation}
\begin{figure}[htb]
    \centering
    \begin{subfigure}[b]{0.45\textwidth}
        \centering
    \fbox{\includegraphics[width=\textwidth]{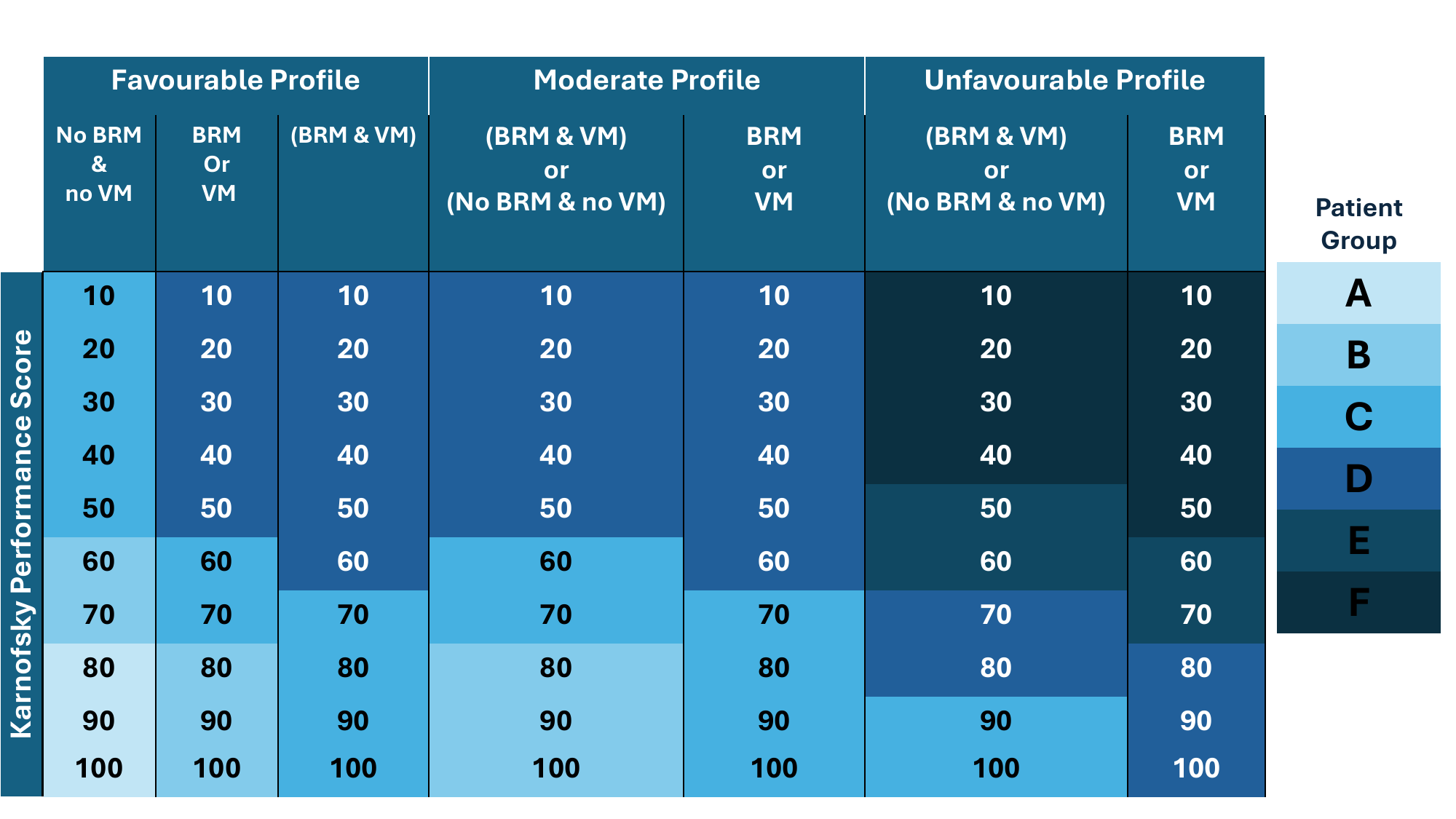}}
        \caption{PISA Cox-based model: Flowchart for patient stratification- \emph{VM} indicates visceral metastases and \emph{BRM} indicates brain metastases.}
        \label{fig:PAL_COX_STRATsubfig1}
    \end{subfigure}   
    \begin{subfigure}[b]{0.45\textwidth}
        \centering
        \includegraphics[width=\textwidth]{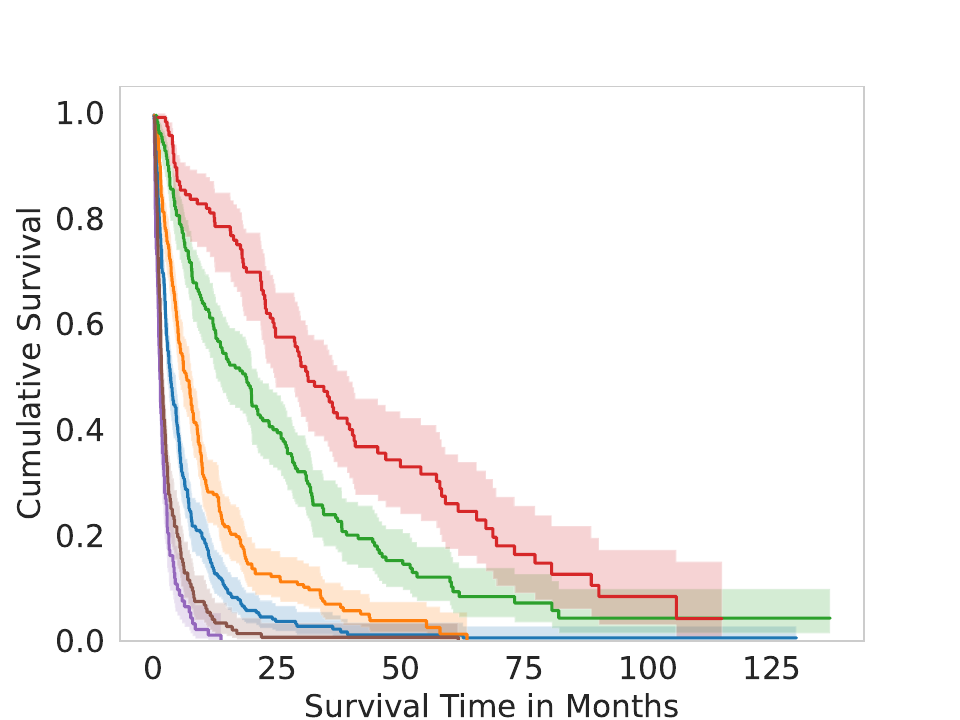} 
        \caption{PISA Cox-based model: Survival curves and 95\% confidence interval on the internal dataset}
        \label{fig:PAL_COX_STRATsubfig2}
    \end{subfigure}

    \begin{subfigure}[b]{0.45\textwidth}
        \centering
        \includegraphics[width=\textwidth]{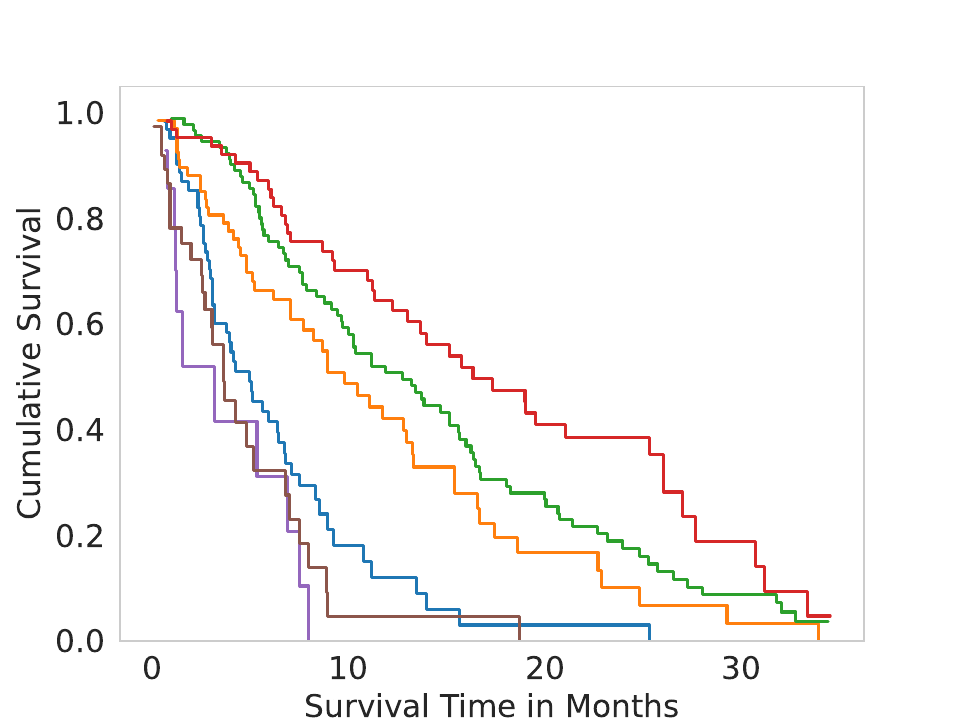} 
        \caption{PISA Cox-based model: Survival curves on the external dataset}
        \label{fig:PAL_COX_STRATsubfig3}
    \end{subfigure}
    \caption{The PISA Stratification Model with Cox regression as the elementary survival analysis model based on Equations \ref{Equation:PAL_COX_1}, \ref{Equation:PAL_COX_2} and \ref{Equation:PAL_COX_3}.
    Subfigure \ref{fig:PAL_COX_STRATsubfig1} shows the flowchart for patient stratification, \ref{fig:PAL_COX_STRATsubfig2} shows survival curves with 95\% confidence intervals on the internal dataset and \ref{fig:PAL_COX_STRATsubfig3} show survival curves on the external datasets. }
    \label{fig:PAL_COX_STRAT}
\end{figure}
Figure \ref{fig:PAL_COX_STRAT} shows the patient stratification based on features $f_1, f_2$ and $f_3$ in Equation \ref{Equation:PAL_COX_1}, \ref{Equation:PAL_COX_2} and \ref{Equation:PAL_COX_3}. This stratification uses 6 groups and achieves an external validation C-index of 0.680. In this model, all original input features are used as in the original study. When analysing the flowchart shown in Figure \ref{fig:PAL_COX_STRATsubfig1}, we observe that, for the favourable profile, survival prognosis is worst with both brain and visceral metastases, slightly better when only one of the two is present, and best when neither is present. For the moderate and unfavourable profiles, survival is less favourable when either brain or visceral metastases is present, but paradoxically improves when both or neither are present. The reason the model does not capture this adequately is that only 3.4\% and 0.05\% (a single patient) of the non-favourable patients in the respective datasets had both brain and visceral metastases. This does, however, emphasise the need for interpretable models, in which clinicians can intuitively check the survival analysis model for such inconsistencies. Because PISA returns multiple models, another model may be picked that does not exhibit this, or, alternatively, a constraint may be introduced that prohibits the observed behaviour explicitly, and then PISA could be re-run.

The survival curves split the patient population into groups, separating both datasets well. Again, the survival curves of both the survival tree and the Cox regression model show that it is challenging to capture the patient groups with very short survival in the internal dataset. 

\subsubsection{Clinical Implications \& Limitations}
By revisiting this previous study, we have shown we can provide a slightly more reliable estimation of survival based on the same original input features available for the previous study, whilst replicating the previous study's main findings. With this, we have demonstrated that it is possible to replace the manual workflow with an automated pipeline. Further, the process can provide additional insight in the relevance of (the interaction of) the different risk factors based on the survival data. Hence, using PISA the process does not have to rely solely on domain expertise.

It is important to note that these datasets were collected more than 15 (and for the external validation set: 25) years ago, with only a limited number of available priginal input features in both datasets. As such, and due to the advancement in medical treatments for certain cancers, the trends found in these models may not be valid for and generalisable to patients with symptomatic spinal metastases today.

\section{PISA Limitations}
There is a trade-off between performance and complexity, and as such, interpretability. To focus on interpretability, we made multiple choices that may restrict model performance. The survival tree depth, the maximum engineered feature size, as well as the maximum number of features are all restricted. Additionally, we choose to restrict the operators available so as to only use simple arithmetic operators. Whilst this ensures that the survival analysis models have enhanced interpretability, some survival analysis use cases may require larger or more complex features/models to maximise performance. For the SUPPORT use case, for instance, the state-of-the-art results were obtained using more complex operators, and, thus, equally good results could not be achieved with our pipeline. If we were to allow equally complex operators, similar results could be obtained (at the cost of higher expression complexity). 

In this work, complexity has been considered a proxy for interpretability. To calculate complexity, the length of an equation is considered, that is, the sum of all operators and original input features used in a feature. A more nuanced, problem or physician-specific, interpretability objective may lead to better trade-offs between performance and interpretability. For instance, if more complex operators were included, an objective could penalise the use of these, favouring the use of simple operators. Additionally, it may be important to limit the number of input features used in the expressions, which could also be considered in an interpretability objective. In general, subjective interpretability choices could be added to an interpretability objective, but these choices would have to be determined prior to running the pipeline. Importantly, however, due to the inherent multi-objective nature of our approach, PISA already returns multiple models that may already contain the models that users are looking for, or alternatively, provide insight into how to change the search. This highlights the benefit of returning multiple models to inspect, rather than a single one.

Further, the use of symbolic regression is a valuable tool to create interpretable models; however, we have used no specific means to prevent symbolic regression from overfitting on the seen data. This problem increases if more complex operators are allowed. Future research could enhance the robustness to overfitting. Moreover, informed choices are made for the GP hyperparameters; however, they are not explicitly optimised for PISA. Optimising these hyperparameters further may yield a performance increase. 

Finally, there are some limitations associated with the computer code for PISA. Specifically, the multiple-feature multi-objective engineering is currently associated with a considerably large runtime, which we note impacts the accessibility of PISA. In future work, this will be combated by enhancing the efficiency via a pure C++ implementation, as well as investigating other avenues that have seemed promising, such as speeding up computations via GPUs~\cite{bouter2018large,bouter2022gpu}. Further, as the usability of PISA is paramount, future work should focus enhancing user interfaces and providing additional guidance to the user on how to select a final model.

\section{Conclusion}

This work introduces PISA - an automated, interpretable survival analysis pipeline that enables clinicians and medical researchers to explore multiple survival analysis models and gain insights into their use case. Beyond model construction, PISA supports patient stratification, that are validated for performance using Kaplan-Meier curves and visualised in flowcharts, all of which are crucial for clinical applicability and understanding. Applied to two clinical benchmark problems, our pipeline achieved state-of-the-art performances, while producing interpretable models and automated patient stratifications summarised in flowcharts that provide additional insights. Revisiting a previous study further demonstrated PISA's capacity to automate and enhance survival analysis workflows in medical research. We therefore believe that PISA has the potential to markedly reduce, if not eliminate, the manual effort traditionally required to create survival models for clinical practice, while delivering interpretable models with state-of-the-art performance.

\FloatBarrier
\section{Methods}
 In this section, the automated interpretable survival analysis pipeline PISA is introduced in detail.
 
The PISA pipeline receives input survival data and a preferred elementary survival analysis method, for instance, Cox regression or survival trees. From this, the pipeline provides a variety of survival models, each of which is built using a different set of features that trade off complexity and performance. In addition, for each survival model provided, the pipeline provides patient risk stratifications as well as the corresponding survival curves. 

Specifically, the pipeline performs several steps, which are outlined below and further detailed in subsequent subsections. 
\begin{enumerate}
    \item \emph{Multiple-feature Multi-objective Feature Engineering} The first step is the feature engineering process, in which multiple feature sets are engineered. Each feature is an expression in which original input features may be combined in a non-linear fashion. In this work, we use feature sets with maximally 3 features to enhance interpretability. The feature sets are constructed so as to trade off complexity (size of the expressions) as well as the performance of the resulting survival analysis model that uses said feature sets. As the feature engineering process is stochastic and survival analysis is very dependent on the exact data used, the feature engineering process is repeated 30 times, each with a different split of the internal set. This typically results in many potential survival analysis models. 
    \item \emph{Model Pre-selection} In the following step, we select the models that provide robust and good performance only. Additionally, further use-case-specific constraints can be added here, such as minimum acceptable performance or specific original input features that must be included. 
    \item \emph{Global and Individual Model Insights} From this, global insights can be generated in the next step. Feature importance scores are provided by observing how many pre-selected models use a certain feature. At this stage, any individual pre-selected model can be investigated and used.
    \item \emph{Patient Risk Stratification} In the final step, we translate all survival analysis models into patient groupings. These groupings are described using flowcharts and analysed via survival curves. Typically, they allow for direct use in clinical practice. 

\end{enumerate}


\subsection{Traditional Survival Analysis Methods}

Although PISA is model agnostic in the sense that any elementary survival analysis model can be used within it, we demonstrate PISA's capabilities in this work with Cox regression and survival tree techniques. These methods are briefly highlighted here before the constituent parts of the PISA pipeline are explained in detail.
\subsubsection{Survival Tree}
Survival trees are typically binary trees that iteratively split the patient population into two groups at each node. These splits are based on a value \(C\) for a patient characteristic \(x\), such that a quality score for the two resulting groups, as defined by the log-rank test score, is maximised~\cite{leblanc1993survivaltree}. That is, the log-rank test score ensures that each node of the tree is split such that the magnitude of separation of the two resulting groups is maximised. Details of the log-rank test can be found in \cite{mantel1966evaluation}.

Finally, survival is estimated using a non-parametric method for each of the tree leaves (final patient groups). 

\subsubsection{Cox proportional hazards regression}
Arguably, the most well-known elementary survival analysis approach is Cox proportional hazards regression \cite{cox1972regression}. Cox proportional hazards regression is semi-parametric. That is, it includes a baseline hazard, which is the same across all patients, as well as a linear log-risk function, also known as the prognostic index, $h(x)=\beta^T*x$, for which beta is determined by maximising the partial likelihood~\cite{katzman2018deepsurv}. Whilst Cox regression is applicable to both uni- and multivariate data, the key limiting assumption is the Cox proportional hazards assumption is met, which assumes that the linear combination of original input features used for risk-modelling remains constant over time \cite{cox1972regression}.

\subsection{PISA Feature Engineering: Multiple-Feature Multi-Objective Symbolic Regression}
\begin{figure*}[t]
    \centering
    \includegraphics[width=0.99\linewidth]{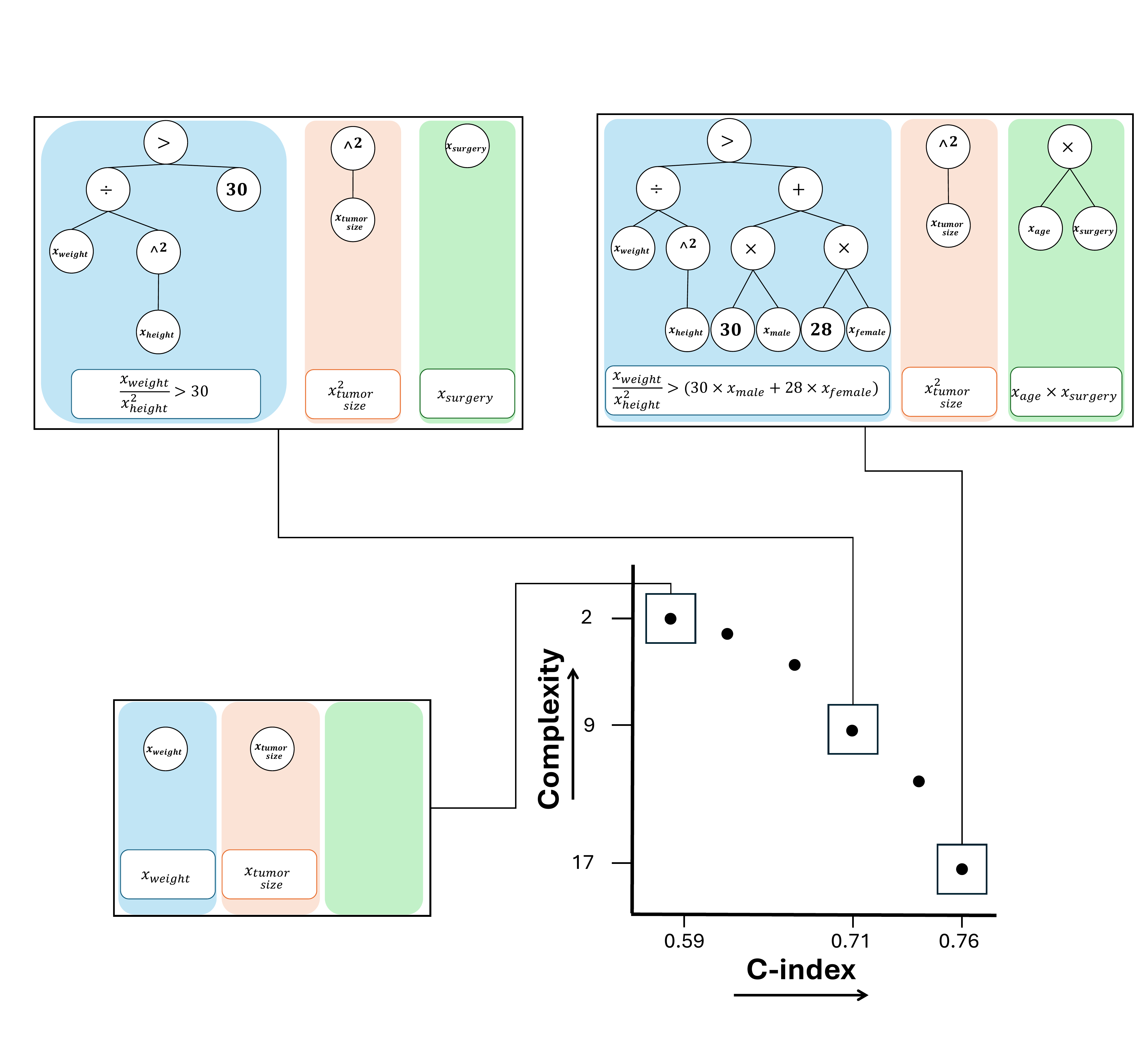}
    \caption{Example of feature sets (with respective expression trees and corresponding feature expressions) returned by the multiple-feature multi-objective GP method, which aims to maximise the C-index performance whilst minimising the complexity. The multiple features (up to $n=3$) within a feature set are shown in different colours and are optimized jointly. Each feature set reflects a trade-off between C-index performance and complexity, and only those sets are kept for which no alternative achieves lower complexity without reducing performance, or higher performance without increasing complexity.}
    \label{fig:MFMO}
\end{figure*}
To engineer features for the survival analysis model, we perform the symbolic regression task. That is, we search for mathematical expressions (features) that describe the risk factors with which survival analysis can be performed well. To do so, we use genetic programming (GP), a technique that can search the space of mathematical expressions spanned by combinations of predefined mathematical operators and input features, structured in the form of an expression tree. 

The specific GP method that we use is GP-GOMEA~\cite{virgolin2020explaining}. The choice to use GP-GOMEA stems from the fact that it is known to have state-of-the-art performance when it comes to finding small, yet accurate expressions, outperforming other forms of GP~\cite{virgolin2020explaining}. This benefits the interpretability of the found features, which is of relevance in (clinical) practice. 

In classic symbolic regression (with GP), a single expression is sought that performs a regression task well. In this work, we use multiple-feature multi-objective GP. GP-GOMEA was previously already adapted for multiple features and multiple objectives by \cite{sijben2022multi} and extended to handle Boolean, categorical and real-valued input features often present in medical datasets by \cite{schlender2024improving}. The multiple-feature aspect means that multiple features in a feature set are generated jointly, i.e., at the same time. This enables finding feature sets in which risk factors complement each other to describe survival well. The multi-objective aspect means that the feature sets are engineered by optimising for two objectives: how well survival modelling can be done using these feature sets, as well as how complex these feature sets are. 
Moreover, search proceeds in a multi-objective fashion. This means that the algorithm searches not for one feature set, but a set of feature sets which exhibit different trade-offs between the two objectives. In Figure \ref{fig:MFMO}, an example of feature sets that could be returned is shown - 
No feature set is returned if it is dominated by another. In this context, one feature set dominates another if it is at least as good in both complexity and C-index performance, and strictly better in at least one of them. In other words, the returned feature sets are those for which no alternative achieves lower complexity without also lowering performance, or higher performance without also increasing complexity.

In the following, we expand on the two objectives before we state the technical hyperparameters used.

\subsubsection{Objectives}
We optimise for two objectives: performance and complexity. 

The performance objective is aimed at assessing how well-suited a feature set is to perform survival analysis with. For this, we build a survival analysis model using the engineered feature set as the only input. No original input features are used, since the aim is to extract and capture relevant risk factors within the engineered feature sets. The performance of a model is evaluated via the Concordance index (C-index)~\cite{harrell1996multivariable}. The C-index is used to assess the quality of the model by considering whether, for each patient pair in a test set, the model predicts the patient with the shorter survival time to have a higher risk.


It is important to note that the reliability of performance measures like the C-index has been criticised by several publications \cite{hartman2023pitfalls,longato2020practical,pencina2015evaluating}. Despite these concerns, the widespread use of the C-index in preceding research and the fact that it enables a straightforward comparison to other state-of-the-art approaches guided our decision to use this metric. We do, however, pay extra attention to the volatility of the C-index to overcome previously flagged issues.
 
Specifically, it is essential to consider the large variance that is inherent to survival analysis. This variance typically comes from a sensitivity to the underlying data distribution. Notably, differences in the numbers of samples, original input features, and censored patients are known to have an impact not only on the C-index, but on other metrics as well, such as the log-transformed, time-independent Brier score~\cite{fernandez2024experimentalcomparisonensemblemethods, prince2025estimation}.  

\begin{figure}
    \centering
    \includegraphics[width=\linewidth]{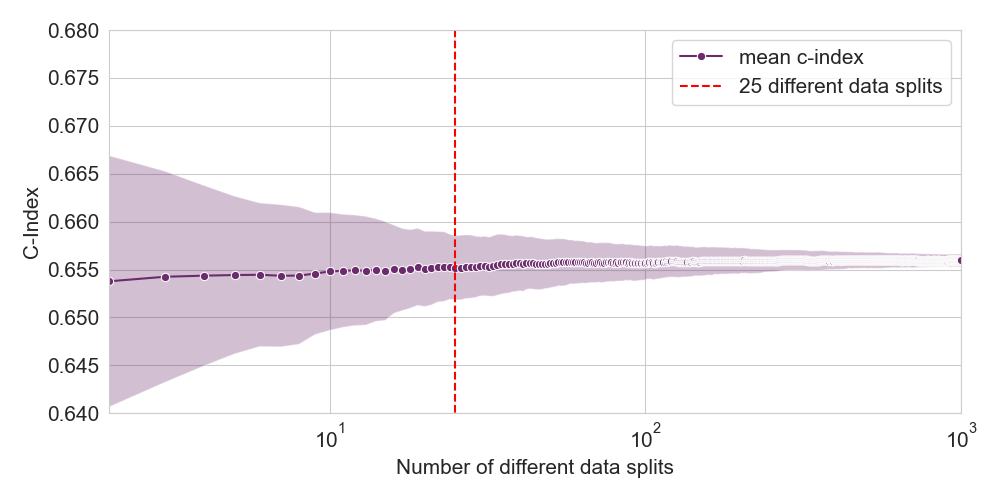} 
    \caption{The mean and standard deviation of the mean C-index of a baseline Cox model for the GBSG dataset computed over different numbers of internal data splits. The mean C-index performance is calculated by building a baseline Cox model and evaluating the C-index on $1$ to $1000$ data splits. This process is repeated 100 times. All data splits are performed by stratifying on the number of patients who experienced an event. }
    \label{fig:Variability_explaines}
\end{figure}

Figure \ref{fig:Variability_explaines} demonstrates this inherent variance, as it shows the variability of the mean C-index over $n$ different data splits of the internal set. Specifically, it shows that the variance of the mean estimated performance decreases the higher number of data splits the mean is calculated over. 

Therefore, to make generalisable statements about a model regardless of the performance metric, survival model, or data distribution, its predictive accuracy should be validated in an unbiased way, using either bootstrapping or cross-validation \cite{harrell1996multivariable}. We do so in this work accordingly, but do make a pragmatic trade-off choice between the increased cost of running multiple models on different data splits and the increased robustness of the performance estimate. Further, instead of using the mean or median to summarise a model's performance over data splits, the interquartile mean is used as it is robust to outlier scores whilst being a better performance indicator as the median~\cite{agarwal2021deep}. The interquartile mean is calculated by disregarding the lowest and highest quartiles of runs and computing the mean on the remaining runs.
Specifically, we define the performance objective to be the interquartile mean of the C-index performances of 25 different train and test set splits (stratified on the censoring rate) of the internal datasets. Figure \ref{fig:Variability_explaines} highlights that 25 different train and test set splits (indicated as the red line) is a good trade-off between variance and run time and should thus give feature sets a representative estimate of their performance. 

For the complexity objective, the size of the expressions is used as a proxy for interpretability, following the rationale that larger features are more difficult to understand. Specifically, the objective is equal to the summation of the size of all features in the feature set that are used in the survival analysis model. Although size is a popular proxy for interpretability, the interpretability of an expression has many factors linked to it, such as the complexity of the mathematical operators used or how many original input features are used within the expression. 

\subsubsection{Hyperparameters}
To maximise the chances of interpretability of our final expressions, we use simple arithmetic and logic operators only. Specifically, the operator set consists of $+,-,\times,\div,x^{2},\leq,=,NOT,AND,OR$. Further, to ensure that the feature set does not grow too large, and, thus, incomprehensible, we flexibly evolve up to 3 features. Each of these features is restricted to include a maximum of 15 operators, original input features and constants.

GP searches for expressions by, for each iteration, making recombinations of existing expressions and evaluating how good these expressions perform. To restrict the run-time of GP, we either stop after 50 of these iterations, or after 5 iterations have the same quality of solutions (defined by a common measure to estimate this quality, the so-called hypervolume~\cite{zitzler2002multiobjective}).

For the elementary survival analysis used within PISA, the implementation and default hyperparameters are taken from Scikit-Survival~\cite{sksurv}. The survival tree depth is limited to a maximum of 3, and the minimum number of samples per leaf is 10. For the Cox regression, we set the regularisation parameter for ridge regression to 1.

\subsection{PISA Model Pre-selection}
The performance of a survival analysis model can vary substantially depending on what
(part of the) internal dataset they were built with. To mitigate this variability and the stochasticity of the feature engineering process, the feature engineering process is applied 30 times. Each application uses a different 75\% of the internal data (internal training data), whereas the other 25\% of the internal data is used as internal validation (internal validation data). To counteract differences in the data distribution, the 75\% internal training and the 25\% internal validation data are sampled so that the number of patients who experienced an event is equally distributed.

To ensure that only models with a stable and clinically reasonable performance are selected for further analysis, we test all solutions as well as the respective elementary baseline model on their respective (unseen) internal validation set. 
We perform 1000 bootstraps on both the internal training and validation sets, whilst ensuring that patients in the internal training set are never used to test the model. The advantage of bootstrapping on both the internal training and validation data to select the models is that this captures the variance of rebuilding the model, rather than just bootstrapping on the internal validation data, i.e. bootstrapping the test set. 
Finally, for each application of the feature engineering process, any model is selected that has no other model's 95\% confidence interval or the corresponding elementary model on the original input features 95\% confidence interval outperforming it.

In practice, the user may have prior knowledge of the survival task or additional constraints that the survival model should satisfy. For instance, a minimally acceptable C-index can be defined here, as well as any required original input features that must be considered. During model pre-selection, these filters can be applied.

\subsection{PISA Model Insights}
A key part of the pipeline is engineering insights for the survival prediction task. For this, we can examine all feature sets that have been found as well as each model individually.

To assess global insights into the survival prediction, the evolved features can be investigated further. To indicate feature importance, we analyse how often each original input feature appears in all the models. The assumption here is that the number of feature sets that a variable is used in correlates with its importance to the survival analysis prediction. Further, we analyse the frequency of sub-expressions occurring for similar reasons. 

Each model at this stage has been found to be of robust quality and is ready to use. Individual model insights are obtained by investigating the engineered features, the survival analysis model, and the resulting patient stratification (described in the next subsection). Specifically, for survival trees, we obtain the limited-depth tree that can be read directly. For Cox regressions, we can observe the hazard ratios of each engineered feature for additional insight. Further, by specifying values for coefficients, the survival curve can be plotted by considering all non-specified covariates to be the respective mean covariate value. With this, it is possible to understand the magnitude and direction of a hazard ratio for a coefficient more intuitively. An example can be seen in Figure \ref{fig:Pipeline}.

\subsection{PISA Patient Stratification}
In clinical settings, the most transparent and widely used method for survival prediction is the use of survival curves, e.g. Kaplan-Meier curves, on relevant patient groups. Kaplan-Meier curves are non-parametric estimators that estimate survival by considering the proportion of patients that are surviving at a current time $t$.  The resulting step function describing the probability of a patient surviving past a given time point is simple and easy to use, but relies on meaningful patient stratification. To this end, we ensure that all models made in our pipeline are transformed into patient stratifications and are summarised in flowcharts. 

The patient stratification leads to groups of patients that receive the same predicted risk, which is defined as the median overall survival of that group in the internal set.

The method of finding the initial patient groups depends on the survival analysis model used and is described below. Once the initial patient groups are found, however, the procedure to promote separability is the same: For each iteration, the pairwise similarity between the patient groups is measured via the log-rank test. If not all groups are determined to be statistically separable with a confidence level of 0.05, the two most similar groups (based on the log-rank) are merged, and a new iteration begins. 

In the following subsections, we elaborate on how these initial groups are found depending on the survival analysis model used.

\subsubsection{Survival Tree}
The advantage of using survival trees is that, by design, the leaves of the tree are non-overlapping groups of patients, which may be considered stratifications. The issue that remains is that two leaves may represent patient groups with the same or similar survival. This is due to the fact that multiple splits in the tree may be needed to fully separate two groups from each other. This can then result in a single group being spread across numerous leaves. However, because we use the merging strategy as identified above, this issue is resolved automatically since if two groups are sufficiently similar, they will be combined.


The benefit of deriving the patient groups directly from the survival trees is that it avoids forcing equal-sized patient groups and, as such, may identify patient groups with extreme prognoses better \cite{royston2013external}. 
\subsubsection{Cox regression}
To generate initial patient groups from a Cox regression model, patients with similar prognostic indices $h(x)$ are grouped. Specifically, we split the prognostic indices into quantiles. To promote the identification of patient groups with extreme prognoses~\cite{royston2013external}, we pay attention to the edge prognostic indices. Specifically, we propose stratifying the prognostic indices into the following quantiles:   $[0.1,0.25,0.5,0.75,0.9,1]$, resulting in 6 initial patient groups. Note that these quantiles can be changed flexibly. 

Although this step already provides patient groups, it is difficult to see which feature values lead to which patient group, i.e. we cannot derive a flowchart. To derive meaningful insights into what feature combinations pertain to which patient group, a decision tree is trained on the non-normalised engineered features to predict the risk group per patient. Using this decision tree, we can derive non-overlapping combinations of features that pertain to a risk group. Before these insights are then translated into the final flowchart, the risk groups are merged with the merging strategy above. 

\subsubsection{Model Agnostic Patient Stratification}
As we propose a pipeline that, in principle, can be used on any survival analysis model, we briefly describe a patient stratification method that uses the engineered features only and does not require any knowledge of the survival analysis model itself.

For this, we split the engineered features into ranges. Each initial patient group is then a combination of a range for each engineered feature.
Specifically, for engineered features with fewer than 7 discrete values, we split the patients into groups based on these values. Whenever the engineered features have more values, the feature is treated as continuous, and patients are split into groups pertaining to 4 quantiles. 

The initial patient groups are then made up of all combinations of value groups for each feature. Finally, the merging process is then applied, as described above.

\bibliographystyle{ACM-Reference-Format}
\bibliography{actual_bib}

\end{document}